%% file: main.tex
\documentclass[nonacm,acmlarge]{acmart}

\usepackage[ruled,linesnumbered]{algorithm2e} 
\usepackage{amsmath}
\usepackage{adjustbox}
\usepackage{booktabs} 
\usepackage{color,soul}
\usepackage{graphicx}
\usepackage{diagbox}
\usepackage{makecell}
\usepackage{multirow,multicol}
\usepackage{textcomp}
\usepackage{setspace}
\usepackage{enumitem}
\usepackage{outlines}
\usepackage{tabularx}
\usepackage{makecell}
\usepackage{colortbl}
\usepackage{caption}
\usepackage{subfigure}
\usepackage{algorithm2e}
\usepackage{array}

\AtBeginDocument{%
  \providecommand\BibTeX{{%
    \normalfont B\kern-0.5em{\scshape i\kern-0.25em b}\kern-0.8em\TeX}}}

\setcopyright{acmlicensed}
\acmJournal{IMWUT}
\acmYear{0} \acmVolume{0} \acmNumber{0} \acmArticle{0} \acmMonth{0} \acmPrice{0}\acmDOI{0}

\begin{document}

\title{Understanding State Social Anxiety in Virtual Social Interactions using Multimodal Wearable Sensing Indicators}

\author{\href{https://orcid.org/0000-0001-5147-3738}{Maria A. Larrazabal}} %
\authornotemark[1]
\email{ml4qf@virginia.edu}
\affiliation{%
  \institution{Department of Psychology, University of Virginia}
  \streetaddress{918 Emmet St N}
  \city{Charlottesville}
  \state{Virginia}
  \country{USA}
  \postcode{22903}
}

\author{\href{https://orcid.org/0000-0002-1611-2053}{Zhiyuan Wang}}
\authornote{Both authors contributed equally to this research.}
\email{vmf9pr@virginia.edu}
\affiliation{%
  \institution{Department of Systems and Information Engineering, University of Virginia}
  \streetaddress{151 Engineer's Way}
  \city{Charlottesville}
  \state{Virginia}
  \country{USA}
  \postcode{22904}
}

\author{\href{https://orcid.org/0000-0003-0705-6704}{Mark Rucker}} %
\email{mr2an@virginia.edu}
\affiliation{%
  \institution{Department of Systems and Information Engineering, University of Virginia}
  \streetaddress{151 Engineer's Way}
  \city{Charlottesville}
  \state{Virginia}
  \country{USA}
  \postcode{22904}
}

\author{\href{https://orcid.org/0000-0002-9152-1391}{Emma R. Toner}} %
\email{ert6g@virginia.edu}
\affiliation{%
  \institution{Department of Psychology, University of Virginia}
  \streetaddress{918 Emmet St N}
  \city{Charlottesville}
  \state{Virginia}
  \country{USA}
  \postcode{22903}
}

\author{\href{https://orcid.org/0000-0001-6295-2523}{Mehdi Boukhechba}}
\email{mmob3f@gmail.com}
\affiliation{%
  \institution{Johnson \& Johnson Innovative Medicine}
  \city{Titusville}
  \state{New Jersey}
  \country{USA}
  \postcode{08560}
}

\author{\href{https://orcid.org/0000-0002-9031-9343}{Bethany A. Teachman}}
\email{bat5x@virginia.edu}
\affiliation{%
  \institution{Department of Psychology, University of Virginia}
  \streetaddress{918 Emmet St N}
  \city{Charlottesville}
  \state{Virginia}
  \country{USA}
  \postcode{22903}
}

\author{\href{https://orcid.org/0000-0001-8224-5164}{Laura E. Barnes}}
\email{lb3dp@virginia.edu}
\affiliation{%
  \institution{Department of Systems and Information Engineering, University of Virginia}
  \streetaddress{151 Engineer's Way}
  \city{Charlottesville}
  \state{Virginia}
  \country{USA}
  \postcode{22904}
}

\renewcommand{\shortauthors}{Larrazabal and Wang, et al.}

\begin{abstract}
Mobile sensing is ubiquitous and offers opportunities to gain insight into state mental health functioning. Detecting state elevations in social anxiety would be especially useful given this phenomenon is highly prevalent and impairing, but often not disclosed. Although anxiety is highly dynamic, fluctuating rapidly over the course of minutes, most work to date has examined anxiety at a scale of hours, days, or longer. 
In the present work, we explore the feasibility of detecting fluctuations in state social anxiety among \textit{N} = 46 undergraduate students with elevated symptoms of trait social anxiety. Participants engaged in two dyadic and two group social interactions via Zoom. We evaluated participants' state anxiety levels as they anticipated, immediately after experiencing, and upon reflecting on each social interaction, spanning a time frame of 2-6 minutes. We collected biobehavioral features (i.e., PPG, EDA, skin temperature, and accelerometer) via Empatica E4 devices as they participated in the varied social contexts (e.g., dyadic vs. group; anticipating vs. experiencing the interaction; experiencing varying levels of social evaluation). We additionally measured their trait mental health functioning. Mixed-effect logistic regression and leave-one-subject-out machine learning modeling indicated biobehavioral features significantly predict state fluctuations in anxiety, though balanced accuracy tended to be modest (59\%). However, our capacity to identify instances of heightened versus low state anxiety significantly increased (with balanced accuracy ranging from 69\% to 84\% across different operationalizations of state anxiety)  when we integrated contextual data alongside trait mental health functioning into our predictive models.. We discuss these and other findings in the context of the broader anxiety detection literature. 

\end{abstract}

\ccsdesc[500]{Human-centered computing~Empirical studies in ubiquitous and mobile computing}
\ccsdesc[500]{Applied computing~Psychology}

\keywords{State social anxiety, passive sensing, wearable sensing, social contexts}

\maketitle

\input{1_introduction}
\input{2_related_work}
\input{3_methodology}

\input{4_results}
\input{5_model_results}

\input{6_discussion}
\input{7_conclusion}

\bibliographystyle{ACM-Reference-Format}
\bibliography{reference,main_aim_2}

\clearpage
\input{Appendix}

\end{document}

%% file: 1_introduction.tex
\section{Introduction}

Mobile sensing is ubiquitous and offers opportunities to gain insight into many important human behaviors, including sleep, activity, productivity, and health. One particularly compelling application of mobile sensing technology involves leveraging it to make inferences about individuals' mental health functioning. Indeed, mental health concerns like anxiety and depression are highly prevalent \cite{bandelow2015epidemiology,kessler2012twelve} and are associated with significant cost to the individual as well as the communities they live in \cite{kessler2013impairment,stuhldreher2014costs}. The dynamic and complex nature of these mental health phenomena calls for measurement methods that can capture them in real time. Evidence to date indicates that mobile sensing biobehavioral data can provide useful insight into individuals' mental health functioning, including negative mood, stress, and anxiety and mood disorders \cite{dobson2023use, place2017behavioral}.

Unobtrusive detection of mental health status may be especially useful to leverage in the case of social anxiety. Social anxiety disorder (SAD) is highly prevalent, with 13\% of adults estimated to experience it at some point in their lifetime \cite{kessler2012twelve}. Moreover, evidence suggests that many college-aged young adults report experiencing severe SAD symptoms (e.g., approximately 40\% as reported by \cite{jefferies2020social}).  Individuals with SAD often report impaired social relationships, workplace functioning, and diminished quality of life \cite{aderka2012functional,barrera2009quality}, and often delay treatment seeking or never seek help because the symptoms of social anxiety (e.g., fear of negative evaluation, self-consciousness) lead them to avoid disclosing their problems. Real-time unobtrusive detection of social anxiety (e.g., via mobile sensing) would facilitate a deeper understanding of the precipitants and time course of anxiety that socially anxious individuals experience, and help clarify the optimal opportunities to offer support. For instance, leveraging smartwatch sensors to better understand how long before entering a known, upcoming social stressor individuals begin to feel anxious, and how long after the stressor before their anxiety returns to baseline would allow us to intervene on this anxiety more effectively (e.g., by encouraging coping efforts during specific times ahead of and following the stressor). Additionally, real-time passive detection of elevated state social anxiety could inform the development of just-in-time adaptive interventions \cite{nahum2018just}, which stand to increase treatment access among this population. Literature to date has indicated the feasibility of leveraging mobile sensing to detect social anxiety \cite{fukazawa2019predicting,jacobson2020digital, jacobson2022digital, stamatis2023specific}, though many open questions remain.

Given the rise of virtual communication options (e.g., FaceTime, Zoom) and their uptake in professional \cite{blanchard2021effects}, educational \cite{labrie2022toward}, and social \cite{bonifati2021holding,bleakley2022bridging} settings, understanding how individuals experience social anxiety in virtual settings seems especially useful. Indeed, there is evidence that socially anxious individuals often prefer online to in-person communication \cite{hutchins2021social}. Moreover, it is plausible that individuals' affective experiences during in-person vs. virtual interactions differ, perhaps in part due to differences between virtual and in-person social interactions (e.g., access to fewer non-verbal cues, which may make it more challenging to `read' their conversation partner). Consistent with this, there is preliminary evidence that in-person interactions, but not virtual modes of communication (e.g., group calls, video calls, voice calling, online messaging, communicating via social media), predict increased positive affect among individuals \cite{liang2023person}.  Further work is needed to gain insight into how individuals experience anxiety during virtual social interactions, given the unique and ubiquitous nature of this mode of communication.    

The present study aims to examine whether we can detect state elevations in anxiety among a sample of undergraduate students with elevated social anxiety symptoms using passive sensing. Specifically, we leverage smartwatch biobehavioral data collected during experimentally-generated social situations on Zoom that were designed to produce variations in state anxiety. This experimental methodology positions us to examine the utility of a range of biobehavioral markers in predicting individuals' self-reported level of social anxiety before, during, and after social interactions that systematically vary on key dimensions (e.g., number of interaction partners). In doing so, the present work makes two central contributions to the growing literature on passive detection of anxiety: 

First, given we possess the ground truth about individuals' social contexts, we are able to examine the benefit of including this contextual information in our classification models. In doing so, we are able to evaluate how biobehavioral features offer insight into in-the-moment anxiety across various controlled social contexts. 
Second, the present work offers insight into how biobehavioral features predict fluctuations in anxiety over brief time scales (a few \textit{minutes}), offering a much more granular view than much existing work (which detects anxiety as it occurs on a daily or weekly level \cite{fukazawa2019predicting,jacobson2020digital,levine2020anxiety}); though see also \cite{jacobson2022digital}. 

%% file: 2_related_work.tex
\section{Related Work}

\subsection{Leveraging Passive Sensing to Detect Self-Reported Anxiety}

A growing body of work has examined the utility of passively sensed biobehavioral features to learn about individuals' self-reported anxiety. Much of this work has examined anxiety symptoms over the course of weeks, and pointed to the value of passive features in predicting anxiety at this level of temporal resolution. For instance, Jacobson \textit{et al.} \cite{jacobson2020digital} predicted the severity of social anxiety symptoms over the past two weeks using accelerometer data and call and text logs from that same time period. They found a correlation of \textit{r} = .7 between symptom levels predicted by the passive features and observed symptoms, with location features playing a particularly important role in the observed relationship \cite{jacobson2020digital}. In another recent study, Stamatis \textit{et al.} \cite{stamatis2023specific} measured social anxiety every 3 weeks throughout a 16-week study, all the while collecting passive information regarding app usage, location, and call and text logs. The researchers then predicted social anxiety symptoms at varying time lags (e.g., 2 weeks in the future, 1 week in the future, and 0 weeks in the future), and found that location and call/text logs were significant and useful predictors of social anxiety symptoms at all time lags. Together, these studies point to the utility of passively sensed information to gain insight into anxiety symptom severity at the level of weeks. 

There is also evidence that passive features can help us detect anxiety at a daily temporal resolution. For instance, researchers have predicted changes in anxiety levels from one day to another using environmental characteristics (e.g., light, sound, location), accelerometer, sleep, and social interaction metrics (e.g., call and text logs, social media use) \cite{fukazawa2019predicting}. Using a combination of features, they were able to predict changes in daily anxiety with 74.2\% accuracy, and  day-of-the-week, brightness, movement, and app use were especially helpful to enhance prediction. Also, Rashid \textit{et al.} \cite{rashid2020predicting} found that passive features, including GPS, pedometer, accelerometer, activity states (e.g., walking, running), and call/text logs predicted daily social anxiety levels. Together, this work suggests that passively sensed biobehavioral features can be successfully leveraged to infer anxious states at a daily level.  

To our knowledge, only one study thus far examined anxiety prediction at the level of hours. Jacobson and Bhattacharya \textit{et al.} \cite{jacobson2022digital} recruited individuals with social anxiety and/or generalized anxiety and had them track their anxiety and avoidance symptoms hourly for one week. Throughout this week, participants' smartphones tracked their physiological activation (heart rate), environmental and location features (e.g., GPS location, light, humidity, temperature), and social contact (i.e., call logs). The results indicated that the information obtained from the passive data helped account for 74\% of the hour-to-hour variance in anxiety and avoidance across the full sample. Idiographic models also pointed to the utility of passive sensing, such that features accounted for (on average) 39\% of the within-person variability in hourly anxiety and avoidance symptoms. While only one study, these findings suggest that passively sensed features hold promise to accurately predict changes in anxiety symptoms, with impressively high temporal resolution (i.e., at the hourly level).  

Taken together, evidence to date points to the utility of passive sensing -- perhaps especially environmental and social interaction features-- to glean information about anxious states over the course of weeks, days, and hours. Extant literature has brought us closer to better mapping anxiety as it occurs "in real time", which stands to benefit not only our understanding of this dynamic phenomenon, but also better position us to assist individuals in coping effectively. However, given anxiety is highly dynamic, and can shift meaningfully in a matter of minutes (vs. hours/days/weeks), it is important to also examine whether passive sensing can assist us in detecting anxiety at this closer temporal resolution. With this in mind, the present study aims to detect shifts in anxiety \textit{over the past 2-6 minutes} using passively sensed smartwatch data. Moreover, although this prior research has examined certain features of individuals' context (e.g., day-of-week, location) and its utility in predicting anxiety, it has not been able to capture details of \textit{the social context} (e.g., the number of people present in the social interaction, the level of social evaluation in the interaction). These social context details stand to strongly influence individuals' experience of in-the-moment anxiety \cite{aldao2013future,hur2020social}, so capturing them may boost our ability to understand anxious states via passive sensing. To this end, the present study leverages the ground truth about experimentally-prompted social interactions to examine the benefits of adding social contextual information to our predictive accuracy. 

\subsection{Leveraging Passive Sensing to Detect Distress in Context}

There is also an accumulating literature evaluating the utility of passive sensing to detect daily distress. Although distress is distinct from anxiety (i.e., it is more general), this literature is relevant, as it provides further insight into the value of passive sensing to infer mental health states. For example, Alinia \textit{et al.} \cite{alinia2021associations} monitored participants' physiology via a smartwatch for 14 days, and also asked them to complete 4 daily surveys. They found that several heart rate variability (HRV) and electrodermal activity features significantly predicted daily levels of stress and negative mood. In a similar study, Boukhechba \textit{et al.} \cite{boukhechba2018demonicsalmon} demonstrated the utility of global positioning system (GPS), accelerometer, and call/text logs to predict daily negative mood in a sample of undergraduate students. Researchers have also reported that location, activity levels, and call/text logs predicted negative affect in the moment \cite{cai2018state}. 

A related body of work has focused on using context-aware and personalized machine learning models to gain deeper insight into distress within specific contexts. Incorporating contextual information into predictive models can serve at least two purposes. First, it can enhance predictive accuracy in identifying and analyzing distress by incorporating contextual factors that might influence such distress \cite{raymond2017modeling}. Second, if certain contexts are found to predict distress, this knowledge can be used to integrate contextual information into timely, context-sensitive digital interventions, i.e., just-in-time adaptive interventions (JITAI; \cite{nahum2018just,coppersmith2022just}). This is especially useful given evidence that failing to capture nuanced user contexts leads to disjointed solutions \cite{heft2001ecological,harari2023understanding}. Work in the ubiquitous and affective computing domain points to the feasibility of this approach and underscores the importance of integrating information about contextual factors for a more comprehensive and accurate interpretation of individual emotional states within various contexts \cite{kanjo2015emotions,poria2017review}. For instance, EmotionSense \cite{rachuri2010emotionsense} is a mobile platform that collects data on users' conversations and context-based information to infer their emotional states in different contexts. Similarly, ``Track Your Happiness'' \cite{killingsworth2010wandering} explores factors affecting happiness in diverse contexts by gathering data on users' feelings and contexts (e.g., at home, at work, in a car) through a smartphone application.


Another avenue we have considered to improve model performance is to integrate users' trait mental health scores, which are typically gathered via brief questionnaires (e.g., the Brief Fear of Negative Evaluation (BFNE) \cite{leary1983brief}) and tend to be comparatively stable over time. This allows models to take into account both relatively stable and more dynamic predictors of mental health state. This integration stands to offer a more nuanced and individualized approach to understanding changing or momentary emotional states. For instance, prior computational psychiatry research \cite{raymond2017modeling} indicates that adding trait anxiety information into computational models predicting anxious states can enhance their relevance and applicability. Specifically, by integrating more stable individual characteristics (e.g., BFNE score), these models can account for more longstanding personal differences in responses to stress or anxiety-provoking situations, resulting in more individualized and accurate predictions.

Expanding upon existing literature, our study examines the distribution of state anxiety across varied contexts and among individuals with varying levels of trait mental health functioning (though all elevated to some degree in social anxiety).  We explore the effect of integrating individual traits and contextual factors in predicting state anxiety across a range of social interactions. 

%% file: 3_methodology.tex
\section{Study Design}

\subsection{Study}

\subsubsection{Participants}
We recruited \textit{N} = 46 undergraduate students who scored high on the Social interaction Anxiety Scale (SIAS; \cite{mattick1998development}), a measure of social anxiety symptom severity. Specifically, individuals who scored at or above 34 on the SIAS were invited to participate, in line with our goal to recruit a sample with moderate-to-severe social anxiety; SIAS scores range between 0-80, and prior work indicates that individuals diagnosed with social anxiety disorder scored 34.6 on this measure, on average \cite{mattick1998development}. 
The average SIAS score in our sample was 45.57 (SD = 8.87), and scores ranged between 34 and 69. 

\subsubsection{Data Sources}
We collected data on individuals' physiological responding and subjective anxiety responses throughout the duration of the virtual study. Additionally, at the end of the study, we asked individuals to report on their trait mental health symptom levels. Each of these data sources is described in turn below. 

\begin{table}[htbp]
\centering
\footnotesize
\caption{State Anxiety Items Measured throughout Virtual Social Interactions}
\label{Table:2}
\begin{tabularx}{\textwidth}{l l l}
\toprule
\textbf{Timepoint} & \textbf{Item Wording} & \textbf{Rating Scale} \\
\midrule
Baseline & I feel... & 1 (\textit{very calm}) to 5 (\textit{very anxious}) \\
Anticipatory & As I think about the upcoming experience, I feel... & 1 (\textit{very calm}) to 5 (\textit{very anxious}) \\
Concurrent & When my feelings were the most intense during the last experience, I felt... & 1 (\textit{very calm}) to 5 (\textit{very anxious}) \\
Post-Event & Looking back on the last experience, I feel... & 1 (\textit{very calm}) to 5 (\textit{very anxious}) \\
\bottomrule
\end{tabularx}
\end{table}

\begin{enumerate}
        \item \textbf{Physiological Responding}. Participants wore an Empatica E4 --a wristworn device-- throughout the study. This device continuously tracked their blood volume pulse via a photoplethysmogram (PPG) sensor, their skin conductance with an electrodermal activity (EDA) sensor, their wrist movement with an accelerometer, and their skin temperature with a temperature sensor. 
        \item \textbf{State Anxiety}. Participants reported on their state subjective anxiety four times throughout each social interaction experience. Specifically, they rated their anxiety once before hearing the instructions about the upcoming experience (i.e., informing them that they would be interacting with one or more other participants; \textit{baseline} anxiety), once after hearing the instructions (\textit{anticipatory} anxiety), once immediately after completing the experience (i.e., having a conversation with one or more other participants; \textit{concurrent} anxiety), and once after reflecting on the experience for 2 minutes (\textit{post-event} anxiety). The wording for each item is provided in Table 1. 

    \item \textbf{Trait Measures. }

    \begin{enumerate}
        \item Social Interaction Anxiety Scale (SIAS; \cite{mattick1998development}). Participants' social anxiety symptom severity was captured using the SIAS. Specifically, participants rated the extent to which each of 20 statements was characteristic of them on a Likert-type scale ranging from 0 (\textit{not at all characteristic of me}) to 4 (\textit{extremely characteristic of me}). Higher scores on the SIAS reflect more severe social anxiety.
        \item Brief Fear of Negative Evaluation (BFNE; \cite{leary1983brief}). The BFNE captures individuals' fear of negative evaluation, which is a core mechanism implicated in the onset and maintenance of social anxiety disorder \cite{heimberg2014cognitive}. Participants indicated the degree to which each of 8 statements characterized them on a 5-point Likert-type scale ranging from 1 (\textit{not at all characteristic of me}) to 5 (\textit{extremely characteristic of me}). Following \cite{rodebaugh2004more}, we administered only the straightforward items. Higher scores indicate greater fear of negative evaluation by others. 
        \item Difficulties in Emotion Regulation Scale - Short Form (DERS-SF; \cite{victor2016validation}). The DERS-SF captures individuals' difficulties with six domains of emotion regulation: accepting emotions, engaging in goal-directed activities when distressed, impulse-control difficulties when distressed, emotional awareness, access to emotion regulation strategies, and emotional clarity. Participants rated how often each of 18 items applied to them on a Likert-type scale ranging from \textit{almost never} (0-10\% of the time) to \textit{almost always} (91-100\% of the time). Emotion regulation is implicated in the maintenance of social anxiety, and individuals with social anxiety have reported difficulties across all domains \cite{helbig2015emotion}.  Higher scores indicate greater difficulties with emotion regulation. 
        \item Depression, Anxiety, and Stress Scales-21 (DASS-21; \cite{henry2005short}). We administered the depression subscale of the DASS-21 to capture individuals' depression symptom severity. Depression symptoms often co-occur with social anxiety \cite{adams2016social}. Participants rated the extent to which each of 7 items applied to them over the past week on a 4-point Likert-type scale ranging from 0 (\textit{did not apply to me at all}) to 3 (\textit{applied to me much or most of the time}). Higher scores indicate higher depression symptom severity. 
    \end{enumerate}

    \end{enumerate}

\subsubsection{Study Procedure}

All study procedures were approved by the Institutional Review Board 
of a large U.S. university and conducted under the supervision of a researcher and licensed clinical psychologist with expertise in anxiety disorders. Researchers delivered the Empatica E4 devices to participants ahead of the study, and participants were scheduled to complete the study in groups of 4-6 on Zoom. Each research session was facilitated by 2-3 undergraduate/graduate researchers. During the study, participants were guided in completing 1 non-social experience (involving watching a non-emotionally evocative video) 
which always happened at the beginning of the study, and subsequently completing two dyadic and two group social experiences (time permitting), which occurred in random order across the study sessions. The social experiences involved having a conversation about a topic provided by the researcher (e.g., "if you had a million dollars, how would you spend it and why?") with one (dyadic) or 3-5 (group) other participants. Two of the social experiences (one dyadic and one group) involved explicit social evaluation (explicitly evaluative experiences), such that participants were told they would be rated by their conversation partners; for these experiences, each participant was asked to rate the extent to which their conversation partner(s) were likeable and good communicator(s) immediately following the conversation. The other two social experiences did not involve this explicit evaluation (though we assume our socially anxious participants still felt some evaluative fears even when no explicit evaluation instructions were given), and are thus termed non-explicitly evaluative experiences. Participants wore the Empatica E4 throughout all experiences, and rated their state anxiety at various points throughout each experience, as described in Section 3.1.2.2. 

\subsection{Data Processing} 

In this study, we analyze wristband biobehavioral data focusing on four primary sensors: photoplethysmography (PPG), electrodermal activity (EDA), accelerometer, and skin temperature. These sensors were selected for their relevance to our research objectives and their alignment with the existing literature on passive detection of anxiety-related states \cite{abd2023wearable}. Specifically, we first segment the data by social experience (Section \ref{sec:data_seg}) and then clean and process the data (Section \ref{sec:data_clean_process}).

\subsubsection{Data Segmentation} \label{sec:data_seg}
The collected data were segmented by research assistants based on the beginning and end points of each phase and experience conducted over Zoom. Each participant's data was divided into multiple segments, corresponding to the different phases of the experiment. This segmentation facilitated a targeted analysis of biobehavioral markers during specific experiences. In this paper, our analysis is confined exclusively to social interactions, including the pre-event, concurrent, and post-event phases of dyadic and group conversations, and excluding all phases of the 'alone video' experience. We opted to focus only on the social experiences given our interest in understanding state anxiety fluctuations as they occur throughout social contexts. 

\subsubsection{Data Cleaning and Preprocessing} \label{sec:data_clean_process}

For both PPG and EDA signals, we applied a 4th order Butterworth filter with specific cut-off frequencies (0.5Hz low cut and 8Hz high cut for PPG, and a high cut of 3Hz for EDA) to remove artifacts and isolate relevant physiological signals. The cleaned PPG signal underwent peak detection to extract systolic peaks and calculate interbeat intervals (IBIs), employing the peak detection algorithm and tools provided by Neurokit2 \footnote{\url{https://neuropsychology.github.io/NeuroKit/}} \cite{Makowski2021neurokit}. For EDA, we further processed the signal to accurately identify the onset and peaks of skin conductance responses, setting an onset threshold at 0.05 Siemens.
We adopted a median filter with a sliding window of 5 seconds on the IBI data to mitigate errors from motion artifacts, followed by a resampling of the median-filtered IBI data from 64Hz to 100Hz using quadratic splines. 

To analyze the IBI power spectral density, we employed the Lomb-Scargle method \cite{fonseca2013lomb}, enhancing our ability to detect and quantify variations in heart rate associated with anxiety states.

Data from the accelerometer and skin temperature sensors were used without further filtering. We used raw sensor readings for deriving meaningful features, such as mean and standard deviation, which are indicative of physical activity levels and thermal regulatory responses during social interactions.

\subsection{Feature Extraction} \label{sec:feature_extract}

From the segmented and cleaned data, features were calculated for each experimental phase per participant. For each experience phase, features were calculated using one-minute non-overlapping time windows, and the features from each window were averaged to obtain the final feature sets. For a broad exploration of potential indicators of state anxiety, we extracted a diverse and rich set of features from the collected biobehavioral, contextual, and trait measure data. 

In addition to this method, we had initially employed an experience length-independent time window analysis, where features were extracted across the full duration of each experimental session, treating it as a single comprehensive window. In this alternative approach, detailed in Appendix \ref{app:feature_setting}, features were extracted across the full duration of each experimental session, treating it as a single comprehensive window. For many (note, not all) of the features, previous research has demonstrated variant window sizes should not significantly impact the analysis \cite{munoz2015validity,shaffer2020critical,baek2015reliability}. However, we ultimately opted to extract features using one-minute non-overlapping time windows because this approach accounts for differences in experience length present in our study design. Notably, though, the analyses in Appendix \ref{app:feature_setting} indicated that the main findings regarding the detection of state anxiety were broadly consistent between the two methods, suggesting robustness in these feature extraction approaches despite variations in window length. 

\subsubsection{Biobehavioral Features} \label{sec:bio_features}

The feature extraction process in this study combined a comprehensive review of biobehavioral markers associated with anxiety with the extensive features that can be extracted by Neurokit2, enabling us to derive a broad set of features.

\textbf{PPG Features:} 
We extracted PPG features across three key domains to provide a comprehensive view of heart rate variability (HRV) and its relevance to anxiety states:

\begin{itemize}
    \item Time Domain: Statistical properties of NN intervals were analyzed to capture the variability in time between successive heartbeats.
    \item Frequency Domain: The distribution of power across various frequency bands of the HRV signal was examined, offering insights into the frequency domain. For the signal transformation from time-domain to frequency-domain, we apply the Lomb-Scargle method to the time-domain samples using evenly spaced basis functions between 0.01 and 0.5Hz \cite{fonseca2013lomb}.
    \item Non-linear Domain: Complexity and non-linear dynamics of the HRV signal were assessed to understand the less predictable aspects of heart rate behavior under stress.
\end{itemize}

\textbf{EDA Features:}
EDA features were derived from both skin conductance response (SCR) metrics and signal-related characteristics:
\begin{itemize}
    \item SCR-related Features: Metrics such as the number of onsets and peaks, height, risetime, recovery, and amplitude were used to quantify emotional and physiological arousal.
    \item EDA Signal-related Features: We analyzed the tonic (baseline arousal and general emotional state) and phasic (rapid conductance changes in response to stimuli) components of the EDA signal, incorporating summary statistics for a detailed understanding of autonomic nervous system activity.
\end{itemize}

\textbf{Accelerometer Features:}
Movement patterns and activity levels were inferred through accelerometer data, including mean and standard deviation of accelerometer signals across x, y, and z dimensions, alongside the calculated magnitude of movement. These measures provide insight into the participant's physical activity and movement dynamics.

\textbf{Skin Temperature Features:}
Mean and standard deviation of skin temperature readings were extracted, offering another dimension of physiological response indicative of emotional and stress states.

\subsubsection{Contextual Features}

This study enriches its feature set by integrating contextual measures alongside physiological data. These contextual measures are coded to capture the nuances of social interactions.  Specifically, the social setting is labeled as 0 for dyadic conversations and 1 for group conversations. The presence of explicit evaluative threat is labeled as 0 for no explicit evaluation and 1 for explicit evaluation. The phase of social interaction is labeled as 1 for the anticipatory phase, 2 for the concurrent phase, and 3 for the post-event phase. This detailed contextual label coding is crucial for understanding how different social environments contribute to state anxiety alongside physiological responses.

\subsubsection{Trait Measures} 

We calculated total scores for each of our trait measures according to the measure's instructions. Specifically, we created a total score for the SIAS by summing all items, for the BFNE by summing all items, for the DERS-SF by taking the mean of all items, and for the DASS-21 depression subscale by summing all items.

\begin{table}[t!]
\centering
\small
\caption{List of extracted features.}
\begin{tabular}{m{2cm} m{12cm}}
\toprule
\emph{\textbf{Category}} & \emph{\textbf{Features}} \\ \midrule
Biobehavioral & \textbf{PPG features:} Interbeat Interval (NN-Interval) Time-domain (Mean NN-Interval, SDNN, RMSSD, SDSD, CVNN, CVSD, MedianNN, MadNN, MCVNN, IQRNN, MinNN, MaxNN, HTI, TINN); Frequency-domain (LF, HF, VHF, LFHF, LFn, HFn, LnHF, SD1); Non-linear measures (CSI, CVI, PIP, IALS, PSS, PAS, GI, SI, AI, PI).\\
& \textbf{EDA features:} Skin Conductance Response (SCR) onsets N, peaks N, height, risetime, recovery, amplitude; Tonic and Phasic EDA signals' median, mean, variance, max, min, skew, kurtosis, standard deviation.\\
& \textbf{Accelerometer features:} Magnitude/x-axis/y-axis/z-axis mean and standard deviation.\\
& \textbf{Skin temperature features:} Skin Temperature mean and standard deviation.
\\ \midrule
Contextual & Social phase category (anticipatory, concurrent, post-event)\\
& Exposure to explicit social evaluation (present vs. absent)\\
& Size of social group (dyadic vs. group)\\ \midrule
Trait Measures & Social Interaction Anxiety (SIAS)\\
& Brief Fear of Negative Evaluation (BFNE)\\
& Difficulties in Emotion Regulation (DERS)\\
& Depression Anxiety Stress (DASS)\\
\bottomrule
\end{tabular}
\label{table:feature-list-updated}
\end{table}

%% file: 4_results.tex
\section{Exploratory Statistical Analysis} \label{sec:statistical_results}

To understand the relationship between state anxiety changes and individual trait measures, contextual factors, and passively sensed physiological markers, we conducted exploratory analyses tied to the following research questions:
\begin{itemize}
    \item[\textbf{R1:}] How are self-reported state anxiety scores distributed sample-wide, within-person, and across individual participants? (Section \ref{section:r1})
    \item[\textbf{R2:}] How are self-reported state anxiety scores distributed across experimentally-manipulated social contexts? (Section \ref{section:r2})
    \item[\textbf{R3:}] How do passively sensed physiological features correlate with self-reported state social anxiety? (Section \ref{section:r3})
\end{itemize}

\subsection{Distributions of State Anxiety Scores at Sample and Individual Levels} \label{section:r1}

\begin{figure*}[t]
\centering
\subfigure[Sample-wide raw state anxiety distribution.]{%
  \includegraphics[width=0.4\columnwidth]{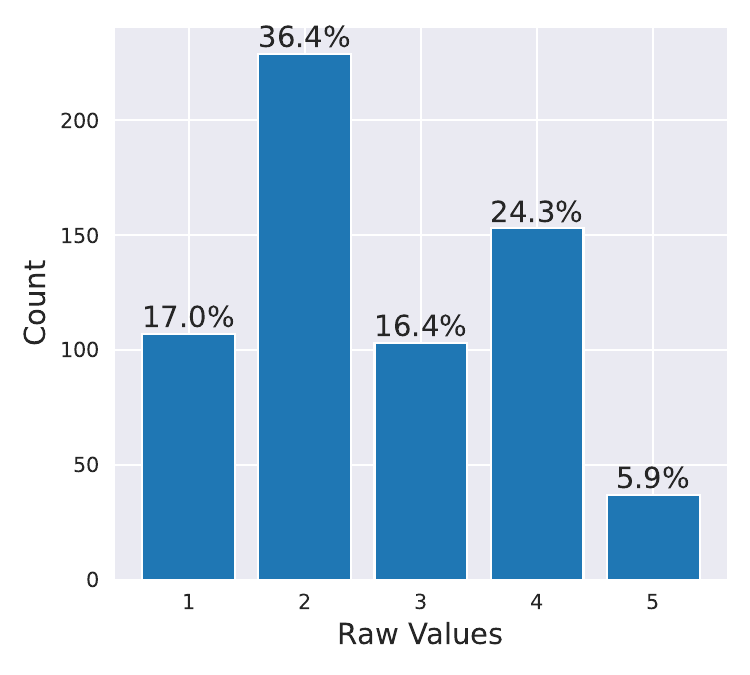}%
  \label{fig:sample_wide_anxiety_raw}%
}
\quad \quad
\subfigure[Adjusted within-person state anxiety distribution.]{%
  \includegraphics[width=0.4\columnwidth]{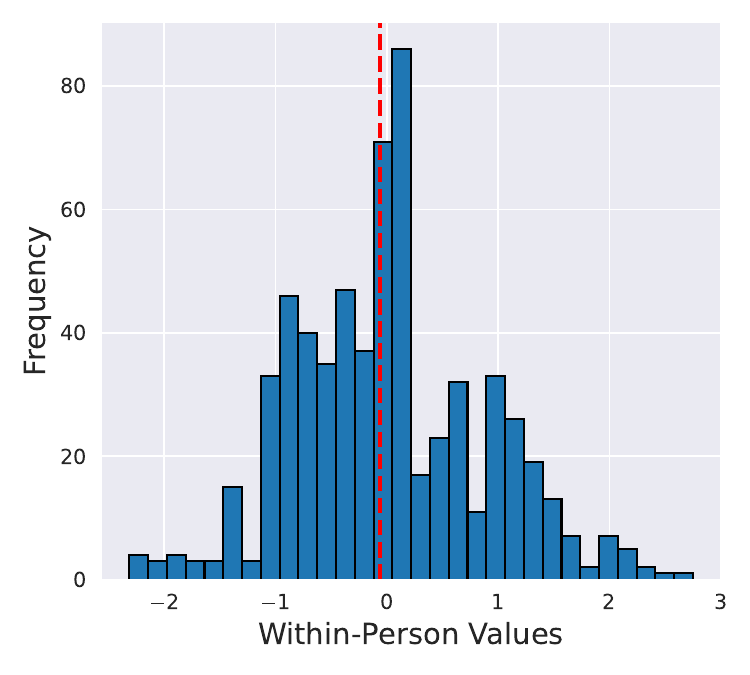}%
  \label{fig:sample_wide_anxiety_within_person}%
}

\subfigure[Distribution Across Individual Participants. (Note: Participant IDs with no associated data reflect individuals who have low trait social anxiety (who are excluded from analyses) or individuals who were scheduled to participate but did not show up to their assigned study participation time.)]{%
  \includegraphics[width=0.85\columnwidth]{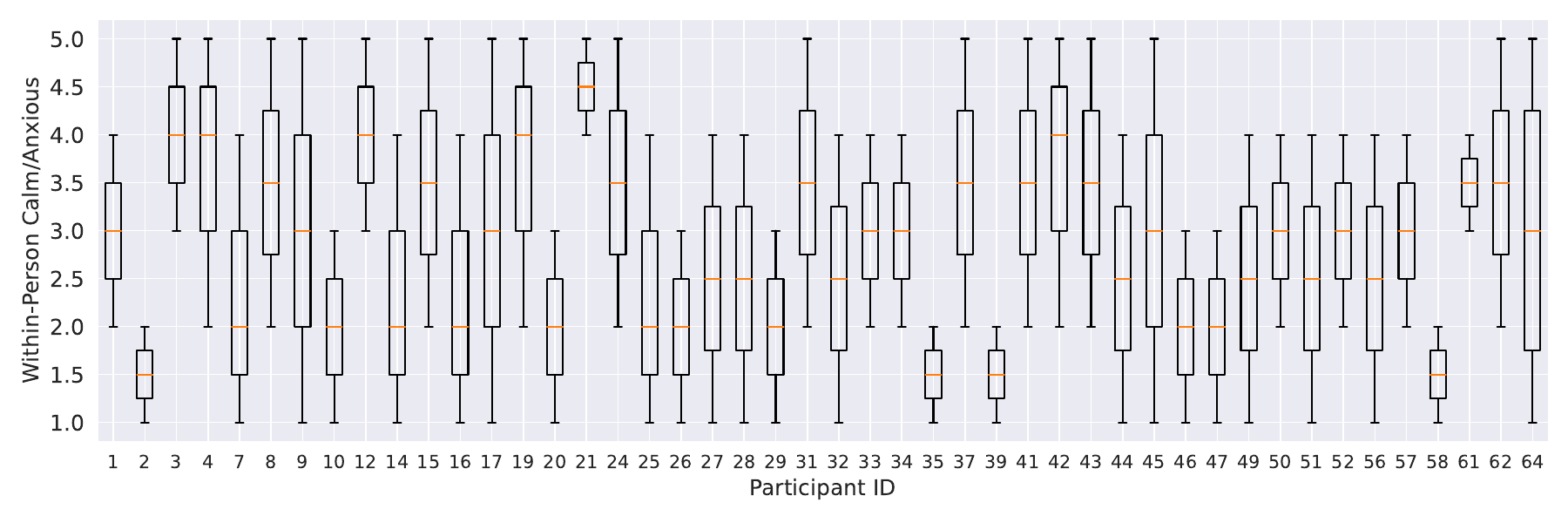}%
  \label{fig:anxiety_across_participants}%
}

\caption{Composite View of State Anxiety Distributions: (a) Sample-wide distribution of raw scores, (b) adjusted within-person score variability, and (c) scores across individual participants.}
\label{fig:sample_wide_anxiety}
\end{figure*}

Below, we provide a summary of  sample-wide general trends in state anxiety within our sample and more nuanced individual differences in state anxiety.

Figure \ref{fig:sample_wide_anxiety_raw} shows participant-reported raw anxiety scores. Most scores fall at the lower end of the scale, followed by responses that gather around the score of 4 and 5, indicating state anxiety is commonly reported. The adjusted within-person scores (the differences between the raw anxiety scores and the mean within-person anxiety scores) in Figure \ref{fig:sample_wide_anxiety_within_person} display a bell-shaped distribution, suggesting that despite variable individual anxiety baselines.

Figure \ref{fig:anxiety_across_participants} provides a boxplot of individual participants' raw anxiety scores, underscoring varied reporting patterns, with notable differences in consistency of high anxiety levels (e.g., P21 vs. P2) and fluctuation across assessments (e.g., P9, 45, and 64). The spread of medians and interquartile ranges highlight the diversity in anxiety experiences within the sample. 

For further insights into the variability of self-reports, particularly focusing on participants with low variability, see Appendix \ref{app:low_varaibility}. Additionally, the proportion of high state anxiety self-reports among participants is detailed in Appendix \ref{app:high_state_anxiety}.

\subsection{Distributions of State Anxiety across Social Contexts} \label{section:r2}

\begin{figure*}[t]
\centering
\subfigure[Across the five experiences. (dyad=dyadic; eval=evaluative; n\_eval=non-evaluative)]{%
  \includegraphics[width=0.48\columnwidth]{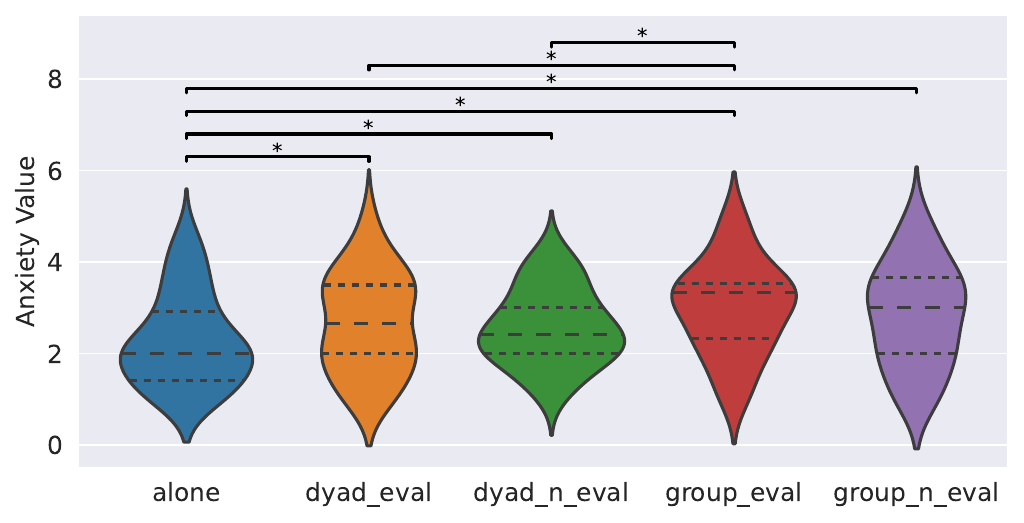}%
  \label{fig:anxiety_across_contexts_experience}%
}
\subfigure[Across temporal phases of anxiety.]{%
  \includegraphics[width=0.48\columnwidth]{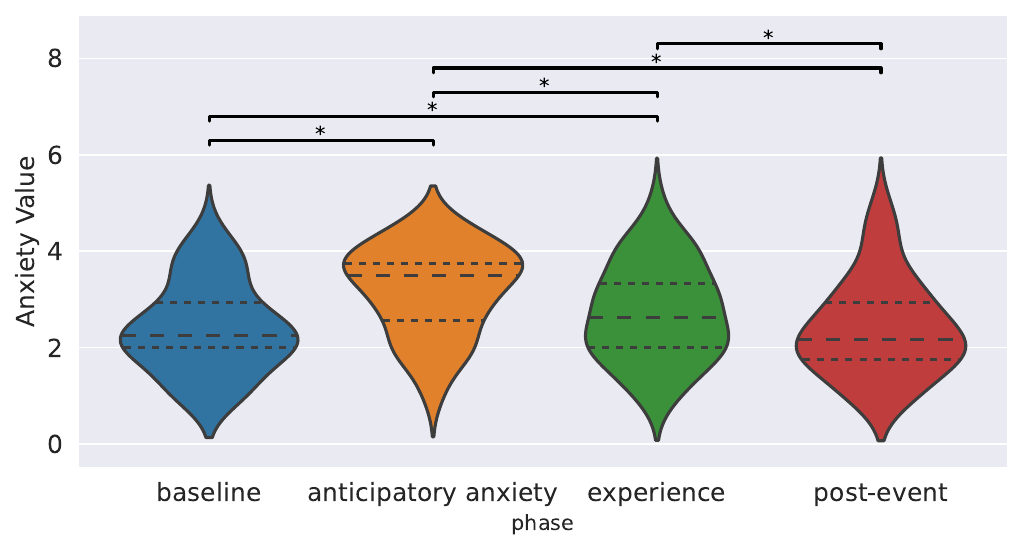}%
  \label{fig:anxiety_across_contexts_phase}%
}
\subfigure[Across sizes of group.]{%
  \includegraphics[width=0.48\columnwidth]{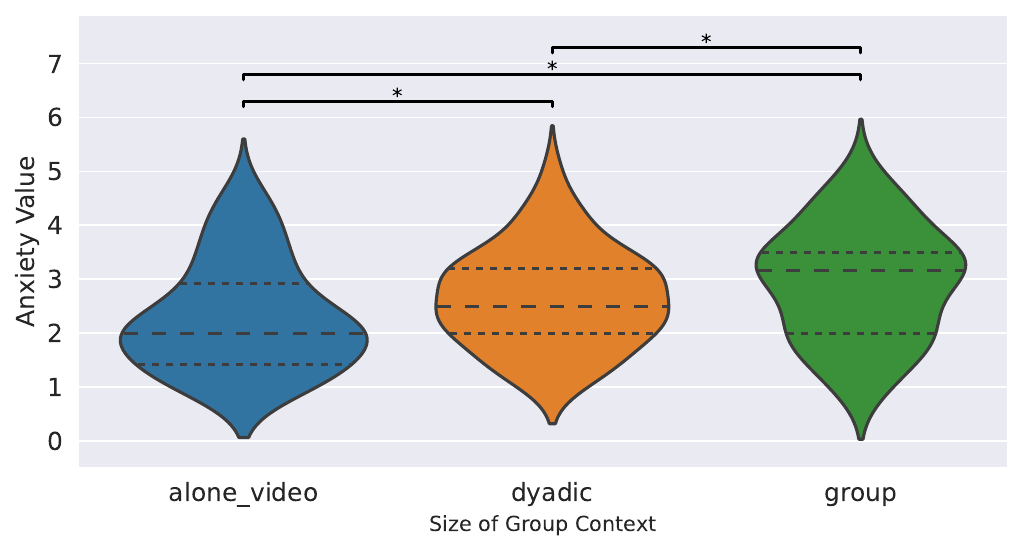}%
  \label{fig:anxiety_across_contexts_group}%
}
\subfigure[Across evaluative threat degrees.]{%
  \includegraphics[width=0.48\columnwidth]{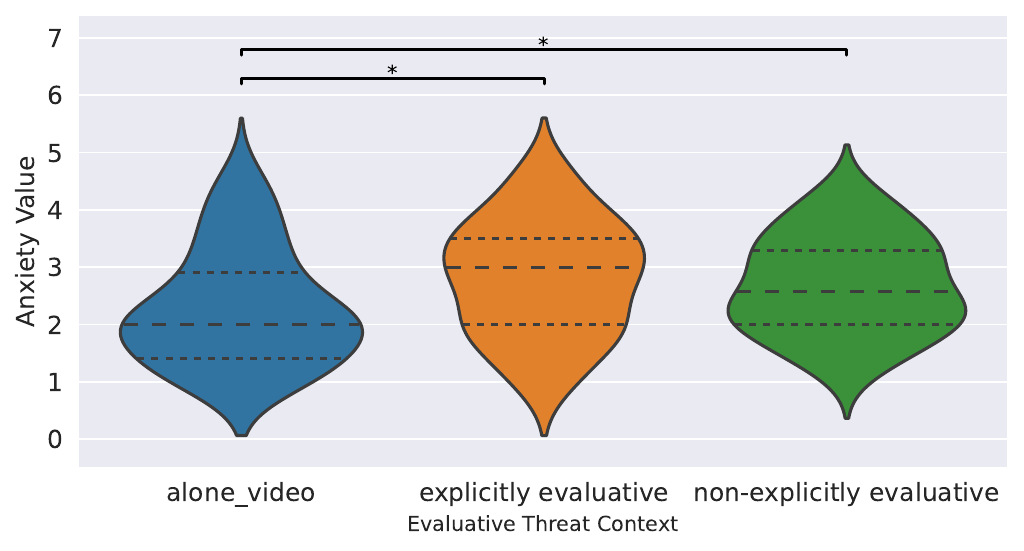}%
  \label{fig:anxiety_across_contexts_eval}%
}
\caption{Comparative Analysis of Anxiety Levels Across Different Social Contexts. This figure presents comparisons of anxiety levels using the paired samples Wilcoxon test across various social experiences (a), phases of interaction (b), group contexts (c), and degrees of evaluative threat (d). The dashed lines within each violent plot delineate the four quartiles. The Benjamini-Hochberg correction was applied to address multiple comparisons within each group of contextual variables in each subfigure, with significance brackets and asterisks marking pairs that demonstrated statistically significant differences at $p<0.05$.
}
\label{fig:anxiety_across_contexts}
\end{figure*}

An examination of the significant test results reveals notable patterns in the data presented in Figure \ref{fig:anxiety_across_contexts}. In Figure \ref{fig:anxiety_across_contexts_experience}, we observe that the anxiety levels are significantly higher in all social experiences compared to being alone. Additionally, the dyadic evaluative context shows significantly different state anxiety scores compared to the group evaluative context. Meanwhile, differences between dyadic evaluative and dyadic non-evaluative, as well as between group evaluative and group non-evaluative, and dyadic evaluative and group non-evaluative are not statistically significant.

In Figure \ref{fig:anxiety_across_contexts_phase}, significant differences are observed among various phases. State anxiety levels during the anticipatory anxiety phase and the experience phase are significantly higher than those in the baseline phase. Furthermore, the anticipatory anxiety phase exhibits the highest anxiety levels, significantly exceeding those in both the experience and post-event phases. During the experience phase, state anxiety scores are notably lower than during the anticipatory phase, yet remain significantly higher compared to the post-event phase.

Figure \ref{fig:anxiety_across_contexts_group} demonstrates that the anxiety levels in group settings are significantly higher than in both alone and dyadic settings, and the dyadic setting also shows a significant increase in anxiety levels compared to being alone.

In Figure \ref{fig:anxiety_across_contexts_eval}, significant findings include both explicitly evaluative and non-explicitly evaluative experiences resulting in significantly higher anxiety levels compared to being alone. However, there is no significant difference between explicitly evaluative and non-explicitly evaluative experiences, despite the differing evaluation instructions across these contexts. Together, these observations advance our understanding of how various social experiences and contexts elicit different anxiety levels, indicating the influence of social contexts on state anxiety.

\subsection{Individual Feature-Level Exploratory Analysis} \label{section:r3}

In this analysis, we examine how state anxiety correlates with individual physiological and contextual features. We employed mixed-effect logistic regression analysis, a statistical method adept at handling data that includes both fixed and random effects, which was important for this dataset, which exhibits both inter-individual variability (random effects) and fixed effects of specific features. The practical purpose here is to control for variability between participants by accommodating random effects, thereby allowing for a more accurate evaluation of how distinct physiological and contextual factors correlate with state anxiety levels. Specifically, the social phase category is delineated into three distinct phases—pre-event anticipatory, concurrent, and post-event—each represented by a one-hot encoding label (e.g., pre-event anticipatory as [1, 0, 0] and concurrent as [0, 1, 0]). By including random effects for each participant, the model discerns the unique influence of each feature on anxiety, beyond individual baseline variations. Prior to modeling, we identified feature outliers as values falling below the first quartile minus 1.5 times the interquartile range (IQR) or above the third quartile plus 1.5 times the IQR. The numeric features were then centered and scaled, which transforms the features by subtracting the mean and dividing by the standard deviation for each feature.

The logistic regression analysis was structured as follows: state anxiety was modeled as a binary outcome (anxious status vs. non-anxious status, with a raw value \(> 3\) indicating anxious status)\footnote{In this paper, we primarily analyzed anxiety status as defined by raw self-reported scores greater than three. For an in-depth exploration of additional analytical methods, including within-person elevations (where state anxiety status is defined by self-reported measures surpassing the individual-level within-person mean) and baseline-based (where state anxiety status is determined by current self-reported state anxiety exceeding the baseline established before the anticipatory anxiety phase) state anxiety definitions, please see Appendix \ref{app:outcome_three}. This appendix provides comprehensive results and discussions on these alternative approaches to understanding state anxiety outcomes.}. Each feature was tested individually against this outcome, with the model structure being:

\begin{equation}
\text{logit}(P(\text{State\_Anxiety} = 1)) = \beta_0 + \beta_1 \times \text{Single\_Feature} + u_{0j} + u_{1j} \times \text{Single\_Feature}
\end{equation}

where \(\beta_0\) is the fixed intercept, \(\beta_1\) the coefficient for \(Single\_Feature\), \(u_{0j}\) the random intercept for baseline variability among participants, and \(u_{1j}\) the random slope representing individual variations in response to \(Single\_Feature\). We measure the statistical significance using p value and Benjamini-Hochberg (B-H) adjusted p-values\footnote{Though we present B-H adjusted p-values here to acknowledge the large number of tests conducted, it is important to note the exploratory nature of this analysis, which aims to identify potentially informative features rather than to draw definitive conclusions, so a correction is not required. To this end, we also highlight the importance of the original p-values.}. 

\begin{table}[t!]
\centering
\small
\caption{Results of Mixed-Effects Logistic Regression Assessing Individual Features' Associations with State Anxiety Self-Reports. Model: State\_Anxiety $\sim$ Single\_Feature + (1 + Single\_Feature|Participant\_ID). Benjamini-Hochberg (B-H) adjustment was applied to the p-values within the feature sets of each independent feature channel (i.e., PPG, EDA, ACC, and contextual features), due to the independence of these sets. Only features with significant raw p-values less than 0.05 are displayed.}

\begin{tabular}{lccccc}
\toprule
\textbf{Feature} & \textbf{Estimate} & \textbf{Std. Error} & \textbf{z value} & \textbf{p value} & \textbf{Adjusted p value} \\ [0.5ex] 
\midrule
\multicolumn{5}{l}{\textit{\textbf{PPG HRV Features}}} \\
Coefficient of Variation of SD (CVSD) & -0.1541 & 0.0017 & -89.5 & <2e-16 & <6.4e-15***\\
Low Frequency Power (LF) & 0.4875 & 0.2129 & 2.290 & 0.0220* & 0.1760 \\
Poincaré Plot Standard Deviation 1 (SD1) & -0.3560 & 0.1694 & -2.101 & 0.0356 & 0.3792 \\
\% NN Intervals in Alternation Segments (PAS) & -0.1762  & 0.0015 & -115.1& <2e-16 & <3.2e-15***\\
\midrule
\multicolumn{5}{l}{\textit{\textbf{Accelerometer Feature}}} \\
Accelerometer Y-axis Standard Deviation &  2.599  & 1.174 & 2.213 & 0.0270 & 0.216\\
\midrule
\multicolumn{5}{l}{\textit{\textbf{Contextual Features}}} \\
Group Size (Dyadic vs. Group)  &  0.7947   &   0.2520  & 3.154 & 0.0016 & 0.0027**\\
Interaction Phase - If Anticipatory & 1.3154 & 0.2491 & 5.281 & 1.29e-07 & 3.23e-07*** \\
Interaction Phase - If Post Event & -1.333  &  0.250 & -5.329 & 9.86e-08 & 4.93e-07*** \\
\bottomrule
\end{tabular}
\label{table:logistic_raw_value}
\end{table}

Table \ref{table:logistic_raw_value} reports statistically significant associations between examined features and state anxiety. The mixed-effect logistic regression analysis highlights significant correlations between various physiological and contextual features with self-reported state anxiety. Among the PPG HRV features, metrics such as CVSD are negatively associated with the likelihood of being in the anxious state, suggesting a decrease in heart rate variability is linked to elevated state anxiety. Additionally, features such as LF are positively associated with elevated state anxiety. The accelerometer feature\footnote{According to \cite{fioretti2020adls}, regarding Empatica E4's accelerometer, the X-axis indicates forward and backward movement along the arm, the Y-axis represents side to side movement across the wrist from the thumb to the pinky, and the Z-axis measures vertical movement up and down from the palm to the back of the hand.} (Standard Deviation of Accelerometer Y-axis readings) also shows a significant positive association before B-H correction, suggesting that greater fluctuations in the lateral wrist movements—as measured by the variability in Y-axis readings—are linked to the highly anxious state. Notably, no EDA and Skin Temperature features were significantly associated with state anxiety. 

Contextual features like the Size of Group (dyadic vs. group) and interaction phases (anticipatory, post-event) reveal significant effects on state anxiety levels. The size feature shows a positive correlation, indicating higher anxiety in larger group settings. This aligns with the sample-wide results presented in Section \ref{section:r2}. The interaction phase features, particularly the anticipatory and post-event phases, display a strong influence on state anxiety levels. The anticipatory phase is positively correlated, suggesting heightened anxiety in anticipation of a social interaction, whereas the post-event phase shows a negative correlation, indicating a possible reduction in anxiety following the completion of the social interaction.

These findings provide critical insights into how individual physiological responses and contextual factors are intertwined with the experience of state anxiety. The variability in these correlations across different features underscores the complexity of the anxiety response and highlights the need for a multifaceted approach in understanding anxiety in context. 

%% file: 5_model_results.tex
\section{Predictive Models of State Anxiety}  \label{sec:machine-learning}

This section examines the feasibility of predicting state anxiety detection. The primary focus is to explore how these models perform in identifying states of social anxiety, and how different factors such as physiological and contextual information influence the effectiveness. The research is guided by the following key questions:

\begin{itemize}
\item \textbf{RQ4:} How well can predictive models accurately identify state anxiety status measured at the level of minutes? (Section \ref{sec:anxiety_detection_performance})
\item \textbf{RQ5:} How does incorporating contextual information and individuals' trait measures enhance the models? (Section \ref{sec:contextual_and_trait})
\item \textbf{RQ6:} Which sensors most add to predictive accuracy, and what is the minimum set of features required for detecting anxiety status? (Section \ref{sec:sensorandfeature})
\item \textbf{RQ7:} How do individual differences in trait anxiety and variability in state anxiety impact the accuracy of machine learning models for detecting state anxiety? (Section \ref{sec:individual_model})
\end{itemize}

\subsection{State Anxiety Detection} \label{sec:anxiety_detection_performance}

In this subsection, we delve into the performance of various machine learning models to accurately identify state anxiety status. The evaluation focuses on the efficacy of these models when incorporating both physiological signals and contextual as well as trait information, with the following evaluation metrics and model configurations:

\begin{itemize}

    \item \textbf{Data Preprocessing}: Our study is designed to test two different approaches to feature engineering for our state anxiety detection models: one approach standardizes features on an individual-level to account for personal baselines and variability, and another one that retains the raw feature values. Furthermore, operationalizations of self-reported anxiety for the model output were also considered, with raw values greater than three indicating heightened anxiety and within-person elevations greater than 0 used to detect deviations from a participant's baseline. 
    \item \textbf{Model Configurations:} The models are configured to either detect general anxiety (versus calm; self-reported score > 3) or identify very high levels of self-reported anxiety (a score of 5). This investigation spans multiple machine learning models generally used in mobile sensing domain \cite{xu2019leveraging,wang2023detecting}, (e.g., Gradient Boosting, Random Forest, and Multilayer Perceptron), each with their model-specific parameters tuned during the validation process. For instance, the maximum depth of a decision tree and the number of neighbors in a K-Nearest Neighbors classifier are optimized to enhance model performance.
    \item \textbf{Model Cross-Validation:} A nested cross-validation methodology was adopted to ensure a robust and unbiased evaluation of the models. An external leave-one-participant-out strategy is coupled with an internal 5-fold cross-validation for hyperparameter tuning. This hierarchical validation scheme is crucial for preventing model overfitting and guarantees that the performance metrics are reflective of the models' true predictive capabilities. Tuned parameters include the number of $K$ best features, which are selected and ranked by the ANOVA F-value between the label and each feature, for training and validation of each individual model. The nested cross-validation ensures that the reported model performances are not artifacts of overfitting to specific participant data or fortuitous parameter selections.
    \item \textbf{Model Performance Metrics}: The effectiveness of the models was measured using two primary metrics: balanced accuracy and macro-F1 score, both set against a random guess baseline of 50\%. The balanced accuracy offers a fair comparison in imbalanced datasets, while the macro-F1 score provides insight into the precision and recall balance of the models.
\end{itemize}

\begin{table}[t!]
\caption{Performance of machine learning models \textbf{with only biobehavioral sensor features} in identifying anxious vs. calm and extreme anxiety.}
\begin{adjustbox}{width=\textwidth,center}
\small
\centering
\begin{tabular}{lcccccccc}
\toprule
\textbf{Self-Reports} & \multicolumn{4}{c}{\textbf{Anxious (Score $>$ 3) vs. Calm}} & \multicolumn{4}{c}{\textbf{Extremely Anxious (Score = 5) vs. Not (Score < 5)}} \\
\cmidrule(lr){2-5} \cmidrule(lr){6-9}
\textbf{Feature Standardize} & \multicolumn{2}{c}{\textbf{Yes}} & \multicolumn{2}{c}{\textbf{No}}
& \multicolumn{2}{c}{\textbf{Yes}} & \multicolumn{2}{c}{\textbf{No}} \\
\cmidrule(lr){2-3} \cmidrule(lr){4-5} \cmidrule(lr){6-7} \cmidrule(lr){8-9}
\textbf{Model} & \textbf{Accuracy} & \textbf{Macro-F1} & \textbf{Accuracy} & \textbf{Macro-F1} & \textbf{Accuracy} & \textbf{Macro-F1} & \textbf{Accuracy} & \textbf{Macro-F1} \\
\midrule
Baseline (Random Guess) & $0.500\pm0.000$ & $0.500\pm0.000$ & $0.500\pm0.000$ & $0.500\pm0.000$ & $0.500\pm0.000$ & $0.500\pm0.000$ & $0.500\pm0.000$ & $0.500\pm0.000$ \\
\midrule
Gradient Boost & \textbf{0.588} $\pm$ \textbf{0.012} & \textbf{0.609} $\pm$ \textbf{0.007} & 0.534 $\pm$ 0.012 & 0.552 $\pm$ 0.008 & 0.812 $\pm$ 0.000 & 0.845 $\pm$ 0.000 & 0.799 $\pm$ 0.003 & 0.846 $\pm$ 0.002 \\
XGBoost & 0.575 $\pm$ 0.000 & 0.593 $\pm$ 0.000 & 0.526 $\pm$ 0.000 & 0.572 $\pm$ 0.000 & 0.815 $\pm$ 0.000 & 0.848 $\pm$ 0.000 & \textbf{0.815} $\pm$ \textbf{0.000} & \textbf{0.858} $\pm$ \textbf{0.000} \\
Random Forest & 0.559 $\pm$ 0.029 & 0.585 $\pm$ 0.015 & 0.565 $\pm$ 0.032 & 0.585 $\pm$ 0.025 & 0.811 $\pm$ 0.002 & 0.855 $\pm$ 0.001 & 0.815 $\pm$ 0.001 & 0.857 $\pm$ 0.001 \\
Decision Tree & 0.562 $\pm$ 0.027 & 0.579 $\pm$ 0.017 & 0.539 $\pm$ 0.011 & 0.560 $\pm$ 0.016 & \textbf{0.819} $\pm$ \textbf{0.009} & \textbf{0.861} $\pm$ \textbf{0.005} & 0.788 $\pm$ 0.008 & 0.839 $\pm$ 0.005 \\
Multilayer Perceptron & 0.552 $\pm$ 0.031 & 0.590 $\pm$ 0.025 & 0.576 $\pm$ 0.037 & 0.579 $\pm$ 0.048 & 0.800 $\pm$ 0.006 & 0.789 $\pm$ 0.003 & 0.814 $\pm$ 0.004 & 0.857 $\pm$ 0.002 \\
Logistic Regression & 0.545 $\pm$ 0.000 & 0.555 $\pm$ 0.000 & 0.582 $\pm$ 0.000 & 0.562 $\pm$ 0.000 & 0.785 $\pm$ 0.000 & 0.831 $\pm$ 0.000 & 0.815 $\pm$ 0.000 & 0.858 $\pm$ 0.000 \\
SVM Classifier & 0.567 $\pm$ 0.000 & 0.562 $\pm$ 0.000 & \textbf{0.583} $\pm$ \textbf{0.000} & \textbf{0.574} $\pm$ \textbf{0.000} & 0.815 $\pm$ 0.000 & 0.822 $\pm$ 0.000 & 0.815 $\pm$ 0.000 & 0.852 $\pm$ 0.000 \\
K-Nearest Neighbors & 0.578 $\pm$ 0.000 & 0.634 $\pm$ 0.000 & 0.514 $\pm$ 0.000 & 0.551 $\pm$ 0.000 & 0.807 $\pm$ 0.000 & 0.832 $\pm$ 0.000 & 0.814 $\pm$ 0.000 & 0.856 $\pm$ 0.000 \\
\bottomrule
\end{tabular}
\end{adjustbox}
\label{tab:results}
\end{table}

\begin{table}[t!]
\caption{Performance of machine learning models \textbf{with biobehavioral, trait, and contextual features} in identifying anxious vs. calm and extreme anxiety.}
\begin{adjustbox}{width=\textwidth,center}
\small
\centering
\begin{tabular}{lcccccccc}
\toprule
\textbf{Self-Reports} & \multicolumn{4}{c}{\textbf{Anxious (Score $>$ 3) vs. Calm}} & \multicolumn{4}{c}{\textbf{Extremely Anxious (Score = 5) vs. Not (Score < 5)}} \\
\cmidrule(lr){2-5} \cmidrule(lr){6-9}
\textbf{Feature Standardize} & \multicolumn{2}{c}{\textbf{Yes}} & \multicolumn{2}{c}{\textbf{No}}
& \multicolumn{2}{c}{\textbf{Yes}} & \multicolumn{2}{c}{\textbf{No}} \\
\cmidrule(lr){2-3} \cmidrule(lr){4-5} \cmidrule(lr){6-7} \cmidrule(lr){8-9}
\textbf{Model} & \textbf{Accuracy} & \textbf{Macro-F1} & \textbf{Accuracy} & \textbf{Macro-F1} & \textbf{Accuracy} & \textbf{Macro-F1} & \textbf{Accuracy} & \textbf{Macro-F1} \\
\midrule
Baseline (Random Guess) & $0.500\pm0.000$ & $0.500\pm0.000$ & $0.500\pm0.000$ & $0.500\pm0.000$ & $0.500\pm0.000$ & $0.500\pm0.000$ & $0.500\pm0.000$ & $0.500\pm0.000$ \\
\midrule
Gradient Boost & $0.674\pm0.005$ & $0.734\pm0.005$ & $0.614\pm0.008$ & $0.652\pm0.008$ & $0.834\pm0.001$ & $0.878\pm0.002$ & $0.831\pm0.003$ & $0.879\pm0.002$ \\
XGBoost & $0.659\pm0.000$ & $0.713\pm0.000$ & $0.636\pm0.000$ & $0.673\pm0.000$ & \textit{0.845}$\pm$\textit{0.000} & \textit{0.888}$\pm$\textit{0.000} & $0.830\pm0.000$ & $0.876\pm0.000$ \\
Random Forest & $0.639\pm0.018$ & $0.682\pm0.024$ & $0.640\pm0.019$ & \textbf{$0.679\pm0.030$} & $0.844\pm0.002$ & $0.886\pm0.001$ & \textbf{$0.843\pm0.005$} & $0.886\pm0.003$ \\
Decision Tree & $0.590\pm0.008$ & $0.637\pm0.013$ & $0.594\pm0.010$ & $0.628\pm0.015$ & $0.833\pm0.005$ & $0.883\pm0.006$ & $0.778\pm0.006$ & $0.844\pm0.004$ \\
Multilayer Perceptron & \textbf{0.695}$\pm$\textbf{0.010} & \textbf{0.716}$\pm$\textbf{0.012}\ & \textbf{0.670}$\pm$\textbf{0.010} & \textbf{0.693}$\pm$\textbf{0.014} & $0.843\pm0.003$ & $0.886\pm0.001$ & $0.842\pm0.003$ & $0.886\pm0.002$ \\
Logistic Regression & \textbf{$0.678\pm0.000$} & $0.710\pm0.000$ & $0.660\pm0.000$ & $0.680\pm0.000$ & $0.839\pm0.000$ & $0.846\pm0.000$ & $0.832\pm0.000$ & \textbf{$0.861\pm0.000$} \\
SVM Classifier & $0.567\pm0.000$ & $0.557\pm0.000$ & $0.583\pm0.000$ & $0.569\pm0.000$ & \textbf{0.845}$\pm$\textbf{0.000} & \textbf{0.888}$\pm$\textbf{0.000} & \textbf{0.845}$\pm$\textbf{0.000} & \textbf{0.888}$\pm$\textbf{0.000} \\
K-Nearest Neighbors & $0.594\pm0.000$ & $0.611\pm0.000$ & $0.592\pm0.000$ & $0.597\pm0.000$ & $0.831\pm0.000$ & $0.863\pm0.000$ & $0.824\pm0.000$ & $0.858\pm0.000$ \\
\bottomrule
\end{tabular}
\end{adjustbox}
\label{tab:results_with_context_trait}
\end{table}

The results, as summarized in Table \ref{tab:results} and Table \ref{tab:results_with_context_trait}, provide valuable insights into the performance of machine learning models in identifying anxious states and extreme anxiety using biobehavioral sensor features alone (Table \ref{tab:results}) and with the addition of trait and contextual features (Table \ref{tab:results_with_context_trait}).

When using only biobehavioral sensor features (Table \ref{tab:results}), the top-performing models, such as Gradient Boost and XGBoost, achieved a balanced accuracy of approximately 58-59\% in identifying anxious vs. calm states indicated by a self-report 1-5 Likert scale > 3. The performance was generally consistent regardless of whether features were standardized by individual or not, with the standardized feature configuration slightly outperforming the non-standardized one. Notably, the models exhibited improved accuracy in detecting extreme anxiety cases (self-report = 5 vs. self-report < 5), with Decision Tree and XGBoost achieving accuracy scores of 81.9\% and 81.5\%, respectively, when features were standardized. Compared to the model performance when identifying anxious (self-report > 3) vs. calm (self-report $\leq$ 3) states, this suggests that the patterns recognized from the features are more pronounced in extremely anxious states (self-report = 5) compared to anxious status (self-report = 4 or 5).

The inclusion of trait and contextual features alongside biobehavioral sensor features (Table \ref{tab:results_with_context_trait}) led to a significant improvement in model performance. The top-performing model, Multilayer Perceptron, achieved accuracies of 69.5\% in identifying anxious (state anxiety > 3) vs. calm states, representing a substantial increase compared to using biobehavioral features alone. The performance was consistent across different feature standardization methods. Furthermore, the models maintained high accuracy in detecting extreme anxiety cases, with several models (XGBoost, Random Forest, Multilayer Perceptron, Logistic Regression, and SVM Classifier) achieving accuracy scores exceeding 84\% across both standardized and non-standardized features.

The improved performance observed with the inclusion of trait and contextual features highlights their importance in enhancing the accuracy of machine learning models for anxiety detection. The consistent performance across different feature standardization methods demonstrates the robustness of the models.

\subsection{Impact of Contextual and Trait Information} \label{sec:contextual_and_trait}

To understand the effect of contextual labels and individuals' trait measures on model performance, we ran ablation experiments by comparing Multilayer Perceptron models (the best performer in Table \ref{tab:results_with_context_trait}) in detecting anxious vs. calm states. The models were trained and evaluated based on biobehavioral features: 1) without contextual and trait information, 2) with only trait information, 3) with only contextual information, and 4) with full information. We also considered two methods for processing self-reported anxiety measures: using raw values and within-person elevations, where each person's mean self-reported state anxiety is used as a threshold to indicate anxiety status.

\begin{figure}[t]
\centering
\includegraphics[width=0.75\textwidth]{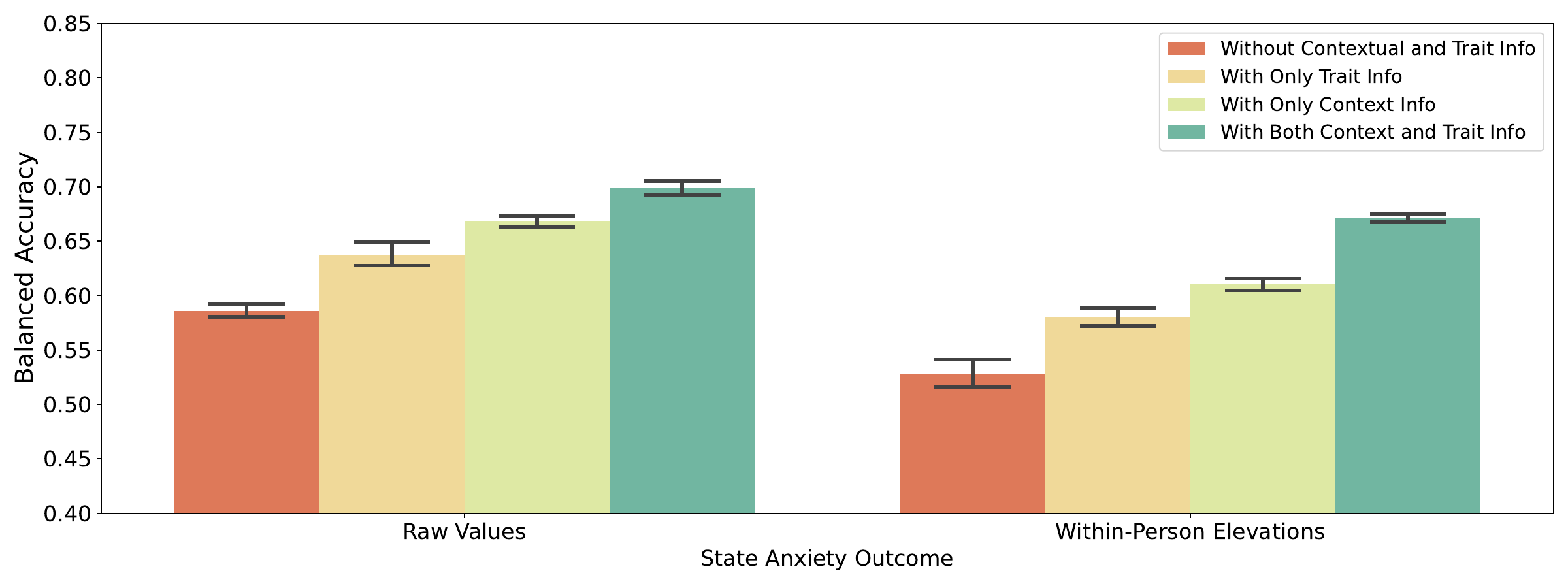}
\caption{Results of ablation experiments examining the impact of contextual and trait measures on model performance alongside biobehavioral features. The error bars quantify the standard deviation of the model performance across 10 training and evaluation experiments.}
\label{fig:ablation_analysis}
\end{figure}

Figure \ref{fig:ablation_analysis} indicates that without incorporating additional contextual information and trait measures, the models can attain an average balanced accuracy of 58.8\% in detecting state anxiety status. Incorporating extra information provides a significant boost in performance, with contextual data increasing accuracy more than trait measures. The inclusion of both contextual and trait information results in the highest accuracy, reaching 69\% for both raw values and within-person elevations. Overall, the inclusion of this additional information results in an accuracy improvement of approximately 10\% compared to using only biobehavioral features. Additionally, comparing the two self-report processing methods, the model accuracy scores were generally higher when using raw values compared to within-person elevations.

\subsection{Sensor- and Feature-Level Importance Analysis} \label{sec:sensorandfeature}

This subsection presents two ablation studies at sensor and feature levels to investigate the importance of each sensor and the efficacy of the features derived from sensor data. Leveraging the models that detect anxiety status by identifying self-reported raw values > 3 or not and using Multilayer Perceptron, we explore the effectiveness of the sensors one by one and features derived from every sensor cumulatively. Specifically, the models were separately trained and evaluated by different subsets of features, in order to compare the importance of sensors and features in identifying anxious status.

\begin{figure*}[t]
\centering
\subfigure[Comparative analysis of model performance across different sensors, with and without contextual and trait information.]{%
  \includegraphics[width=0.49\columnwidth]{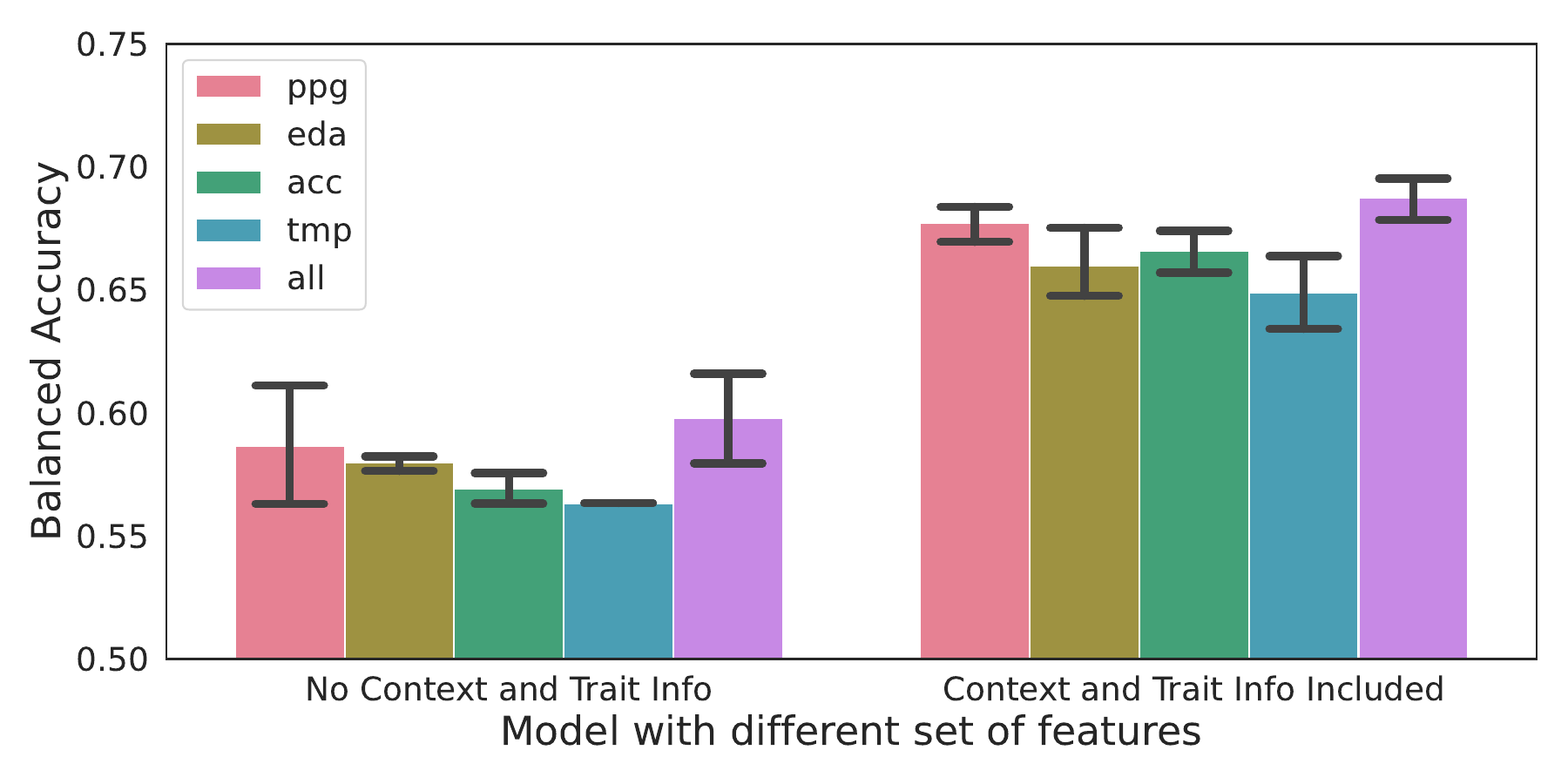}%
  \label{fig:sensor_level}%
}
\subfigure[Model performance as a function of the number of top features selected from individual sensors alongside contextual and trait information.]{%
  \includegraphics[width=0.5\columnwidth]{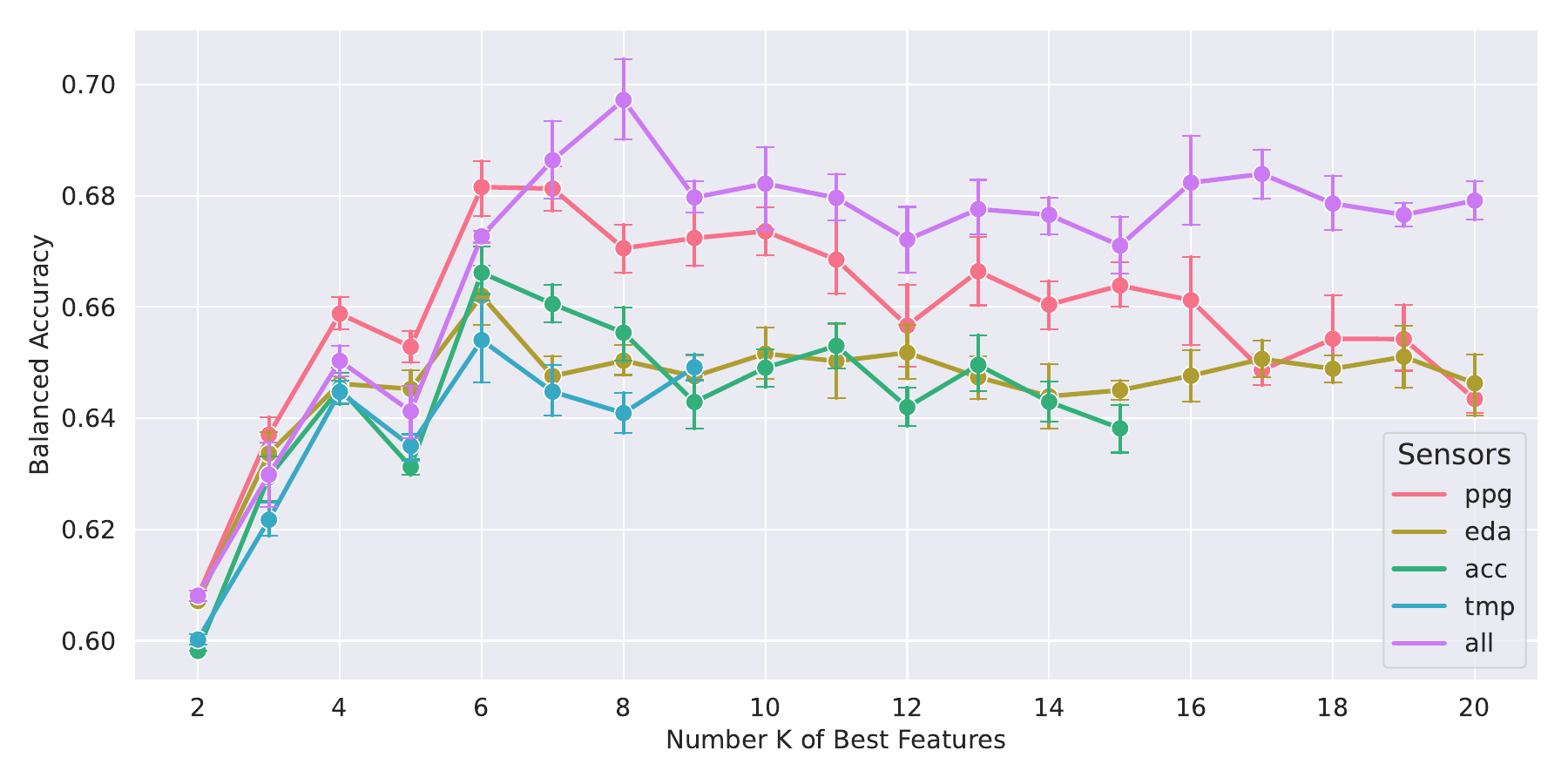}%
  \label{fig:feature_level}%
}
\caption{Comparative analysis of model performance: sensor-level contributions (left) and feature-level significance within sensors (right) in ablation studies. The error bars in both figures represent the variability in model performance across different model iterations for each participant individually.}
\label{fig:sensor_effectiveness}
\end{figure*}

\textbf{Sensor-Level Analysis:} As shown in Figure \ref{fig:sensor_level}, the analysis examines the influence of various sensors on the predictive outcomes, considering scenarios both with and without contextual information. The PPG sensor stands out for its substantial impact, while the EDA sensor contributes relatively less to state anxiety detection. The accelerometer sensor's influence is moderate when isolated. However, when contextual and trait information is introduced, the accelerometer sensor's effectiveness increases comparably to other sensors, making it the second strongest sensor, surpassing the EDA sensor. In contrast, the skin temperature sensor consistently contributes the least to predictive accuracy out of all sensors, regardless of the presence of contextual and trait information. However, the introduction of contextual and trait information markedly enhances model performance for all sensors, with significant improvements observed for each sensor configuration. This suggests that the patterns detected by these sensors become more predictive of state anxiety when contextualized.

\textbf{Feature-Level Analysis:} In our detailed feature-level analysis (see Figure \ref{fig:feature_level}), when considering each sensor modality alongside contextual and trait information, the model's behavior is connected to the choice of the top `$K$' features, emphasizing the significance of specific features. Notably, the PPG sensor showcases exceptional efficiency; it achieves a balanced accuracy of over 68\% with only six top features, which is remarkably close to the nearly 70\% peak accuracy attained using the top eight features from a combination of all sensors. 

We also manually checked the most frequently selected top features across the points indicated in Figure \ref{fig:feature_level}. Notably, under the configuration using the top eight features from all features, which contributed to the peak performance, the most frequently selected eight features across the LOOCV rounds are: Social Phase ID (anticipatory vs. concurrent vs. post-event), SIAS measure, PPG Standard Deviation of the Successive Differences (SDSD), BFNE measure, Number of EDA SCR Peaks, DASS measure, PPG Root Mean Square of Successive RR Interval Differences (RMSSD), PPG Normalized High Frequency Ratio (HFn), indicating the important roles of physiological, contextual, and trait information to optimize anxiety detection outcomes.

\subsection{Individual-Level Model Performance Analysis} \label{sec:individual_model}

Under Leave-One-Out Cross Validation (LOOCV) strategy, we also examine the relationship between model performance for every individual model and individual participants' social anxiety at both trait and state levels. For a comprehensive understanding, each participant's model was trained and validated 10 times to obtain averaged balanced accuracy scores of the individual models. On this basis, we investigate the relationship across individual model performance (i.e., balanced accuracy), individuals' trait anxiety (SIAS), and self-reported state anxiety measures across the experimental sessions (i.e., their means and standard deviations).  

\begin{figure}[t]
\centering
\includegraphics[width=0.7\textwidth]{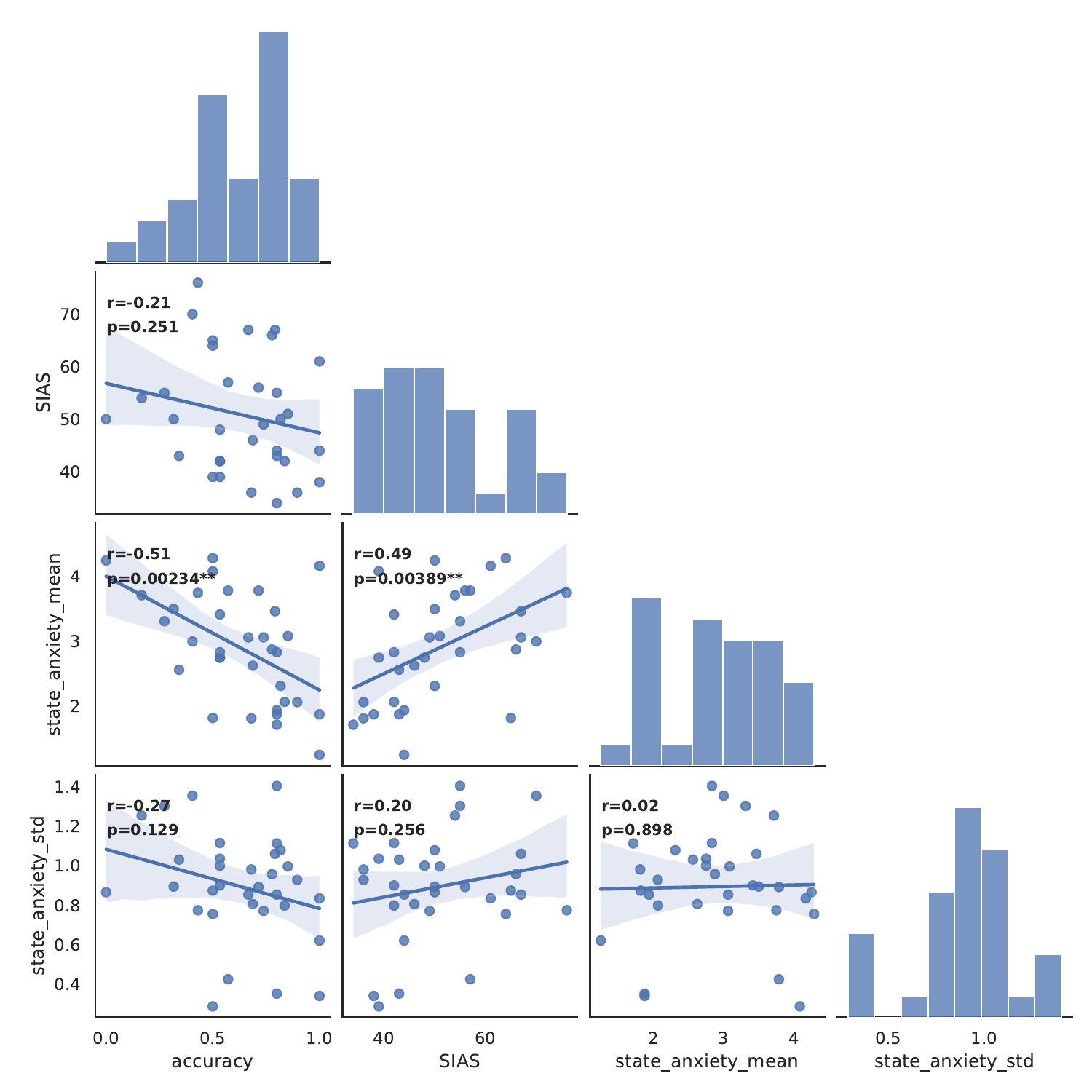}
\caption{Scatter plots with Pearson-correlation lines illustrating the relationship between sample-wide individual-level model accuracies, trait anxiety SIAS scores, and means and standard deviations of state anxiety.}
\label{fig:individual_level_model_performance}
\end{figure}

Figure \ref{fig:individual_level_model_performance} reveals statistically significant Pearson correlations among two pairs of measures. First, higher trait anxiety SIAS scores predicted higher mean state anxiety scores (\textit{r}= .49, p=0.00389**), suggesting that individuals who reported higher trait social anxiety symptoms also reported higher average state anxiety scores during the social experiences, as would be expected. Second, higher model accuracy scores predicted lower mean state anxiety scores (\textit{r}=- .51, p=0.00234**), suggesting that accuracy was greater when individuals tended to report lower state anxiety, on average. 

We also observed negative correlations between: 1) SIAS scores and individual model accuracies (\textit{r}=- .21, \textit{p}= .251), and 2) individual model accuracy scores and state anxiety standard deviations (\textit{r}= -.27, \textit{p} = .129), as well as a positive correlation between SIAS scores and state anxiety standard deviations (\textit{r} = .20, \textit{p} = .256).


%% file: 6_discussion.tex
\section{Discussion}
In this study, we used an experimental paradigm to examine fluctuations in state social anxiety throughout several manipulated dyadic and group social interactions. As we discuss below, our findings shed light on the dynamic nature of state anxiety, its manifestation across various social contexts, and the potential utility of machine learning models to recognize anxiety states with notable precision. Moreover, our findings point to the value of incorporating contextual and trait information to improve model performance. 

\subsection{State Social Anxiety was Highly Variable Between and Within People}

Our investigation into the distribution of self-reported state anxiety revealed pronounced variability across participants in our sample. Specifically, some participants consistently reported elevated anxiety levels, particularly if they reported higher trait social anxiety symptom severity. Meanwhile, other participants exhibited significant fluctuations across low and high state anxiety levels, regardless of their trait social anxiety levels. This variability highlights the complex and individualized nature of anxiety experiences and is in line with prior research indicating high variability in state negative emotions among socially anxious individuals \cite{farmer2014affective,kashdan2014differentiating}. Indeed, patterns of within-person fluctuations in state anxiety revealed a close-to-normal distribution. 

Examining state anxiety across different social contexts pointed to factors that may have contributed to the observed variations. For instance, participants experienced significantly higher state anxiety levels when in social settings compared to when alone, suggesting our social interactions effectively provoked temporary social anxiety. Additionally, individuals reported the highest state anxiety scores during the anticipatory phase (i.e., a 2-minute period during which they were instructed to think about the conversation they were about to have with another participant(s)) relative to when they were interacting with others in the study.  This is consistent with evidence that highly socially anxious individuals experience elevated anxiety when anticipating a social interaction (i.e., anticipatory anxiety; \cite{sluis2017repetitive,vassilopoulos2004anticipatory}. Their state anxiety decreased even further after the interaction was over (i.e., post-event phase), which may reflect relief from finishing an anxiety-provoking social interaction. We also found that individuals reported significantly higher state anxiety in larger interaction groups (4-6 people) than during dyadic conversations, suggesting individuals found trying to navigate multiple social interaction partners simultaneously especially anxiety-provoking. Meanwhile, experimentally-manipulated evaluative threat levels did not predict variations in state anxiety levels, suggesting social interactions were experienced as equally anxiety-provoking regardless of whether evaluation was made explicit or not, perhaps because evaluative concerns were inherently activated by the social situations for our anxious sample even when not made explicit. Together, these findings indicate that, in our sample, individuals tended to experience the most anxiety if they reported high trait social anxiety symptoms, and when anticipating social interactions, especially group-based ones. 

\subsection{Biobehavioral Features Alone Predict State Social Anxiety}

Models leveraging biobehavioral features alone were able to predict elevations in state social anxiety at above-chance levels, and were especially strong at predicting when elevations were `extreme' (i.e., anxiety scores of 5). That is, using biobehavioral features from our four sensing streams, models detected low vs. high state anxiety levels with balanced accuracy scores around 58-59\%. Although this accuracy level is better than random guessing, it definitely leaves room for improvement. This relatively low predictive accuracy is in line with evidence that individuals reporting different levels of state anxiety exhibit similar physiological responding \cite{mauss2004there}, suggesting other factors beyond physiological responding will be key to differentiating small changes in mild to moderate state anxiety. Interestingly, however, our models did quite well at predicting `extreme' elevations in state social anxiety (i.e., scores of 5), displaying balanced accuracy levels around 81\%. This indicates that physiological markers can inform us about state anxiety fluctuations, though perhaps only when elevations are strong enough in magnitude. 

Statistical analyses coupled with sensor-level and feature importance analyses provided further insight into the utility of biobehavioral markers in state social anxiety prediction. Across the board, PPG emerged as an important predictor of state social anxiety variations. For instance, mixed effects logistic regression analyses indicated that four PPG HRV features were statistically significant predictors of the likelihood of endorsing high vs. low state social anxiety. Specifically, lower HRV (indicated by CVSD, SD1, and PAS) and a higher low frequency power (LF) component of HRV are both associated with a greater likelihood of endorsing elevated state anxiety. This is consistent with prior evidence linking HRV to social anxiety symptoms \cite{alvares2013reduced, cheng2022heart} and state-level distress \cite{alinia2021associations}. Meanwhile, only $1$ accelerometer feature, and no skin temperature or EDA features, were significantly associated with elevated state social anxiety status. Additionally, sensor importance analyses indicated that, when considering each sensing stream individually, PPG features offered the highest balanced accuracy level, which was comparable to the balanced accuracy level achieved when using all sensing streams. Moreover, feature importance analyses revealed that PPG features were highly efficient at predicting state social anxiety status. Models detecting state anxiety status achieved 68\% accuracy by using only the top 6 features (relying only on PPG features alongside contextual and trait features), which is very close to the 70\% accuracy level reached when using the top 8 features with all sensing modalities leveraged. Together, results indicate that PPG as a sensor and PPG features may be very useful to our prediction of state social anxiety elevations. This is in line with prior research linking PPG HRV features with anxiety \cite{kim2018stress}. Although other sensing streams (e.g., accelerometer) also offered some utility (in line with \cite{boukhechba2018demonicsalmon,fukazawa2019predicting,jacobson2020digital}, PPG consistently emerged as a top predictor across our different analytic approaches. 

\subsection{Adding Contextual and Trait Information Boosted Predictive Accuracy}

Our ability to successfully predict state social anxiety was stronger when we incorporated information about social context and trait mental health functioning. For instance, adding information about the number of interaction partners (dyad vs. group), the temporal phase (anticipatory, concurrent, post-event), and the level of evaluation (explicit vs. not), as well as adding trait social anxiety, depression, and emotion regulation scores boosted our model's predictive capability from ~58\% to 69.5\%. That is, adding this information improved our ability to predict whether individuals reported low vs. high (scores >3) by 10\%. Adding context and trait information to models also improved prediction of extreme anxiety from 81\% to 84\%, pointing to the utility of these information sources for predicting extreme anxiety elevations. 

Results further indicated that contextual information may be especially useful (relative to trait information) to state anxiety prediction efforts. Adding contextual information alone to the model boosted performance to 67.12\%. This is consistent with prior work indicating that a combination of contextual (e.g., location, weather, time) and physiological (i.e., HRV) features help predict hour-to-hour changes in anxiety \cite{jacobson2022digital}. While adding trait information alone also benefited the model, the benefits were not as great as those associated with adding context information alone. Nonetheless, it is important to highlight that the highest model performance occurred when we added \textit{both} context \textit{and} trait information on top of biobehavioral features, pointing to the utility of all three information sources for state social anxiety detection. Indeed, analysis of the top features indicated that social anxiety and depression symptoms, PPG features, and temporal phase of social interactions made the most useful contributions to our prediction. This highlights that all sources of information are useful to state anxiety prediction efforts. 

Interestingly, we found that our classification models were more accurate among individuals who reported lower mean state social anxiety. One potential explanation for this is that these individuals tended to consistently report lower anxiety, making their rare elevations in state anxiety easier to detect because they clearly stood out as distinct from their normative low anxiety response pattern. However, this seems insufficient as an explanation given that accuracy was unrelated to the amount of variability (i.e., SD) in state social anxiety. Further work is needed to clarify the link between state social anxiety levels and accuracy. Additionally, it is notable that our models were equally accurate regardless of levels of trait social anxiety severity and variability in state social anxiety. This suggests our models may function well across a wide range of anxious response tendencies. 

\subsection{Implications for State Anxiety Detection}

Our findings point to at least three implications for state anxiety detection work. First, our results consistently indicated that contextual and trait information benefits our ability to detect state social anxiety elevations. This was true for both state anxiety cutoffs we used, though especially for our primary state anxiety cutoff (i.e., scores >3 vs. not). This suggests that, to the extent that it is feasible, researchers may find it worthwhile to collect data on contextual characteristics and/or trait mental health functioning. The more we can make collecting and integrating this data low burden for researchers, clinical providers, and anxious individual, the more scalable and meaningful these approaches will be.

Second, we found that our ability to detect state social anxiety was much better when focusing on `extreme' state elevations (i.e., whether the maximum anxiety score was endorsed or not). This was true for our models including only biobehavioral features as well as for those incorporating context and trait information. This pattern may reflect that increased physiological activation was mostly occurring when individuals endorsed feeling extremely state anxious (i.e., more of a threshold than continuous model), and therefore our models were able to differentiate these extreme elevations more readily. Although further work is needed to better understand this pattern, our findings suggest that testing different thresholding approaches for state social anxiety may pay off, as it may reveal ways to binarize state social anxiety that offer better predictive capabilities. 

Third, our results indicated that including more sensing streams and features in our models is not always necessary or beneficial. Indeed, our sensor importance analyses revealed that including PPG alone afforded similar levels of predictive accuracy as including PPG, EDA, skin temperature, and accelerometer combined. Although these results are preliminary, if replicated, they would suggest that researchers seeking to predict state anxiety can prioritize PPG sensors as a way to cut down on resources needed and increase efficiency. Additionally, our feature importance analyses indicated that, for single-sensor analyses, accuracy often peaked at ~6 features, and leveled off after that, and for analyses including all sensors, accuracy peaked at 8 features, and decreased after that. Therefore, including a greater number of features not only did not benefit the model, but in some cases, was associated with poorer performance. This indicates that researchers may benefit from incorporating approaches that identify the top \textit{k} features as they build their classification models. 

\subsection{Limitations and Future Work}

Our findings should be interpreted in light of several limitations. First, our study experimentally manipulated social contexts on Zoom; while we believe examining virtual social interactions is a critical area of study given their ubiquity, this does mean our findings may not generalize to naturalistic virtual interactions and in-person interactions. Further work is needed to examine whether these findings emerge in other, less tightly controlled and/or in-person settings. Second, although our study extends upon prior work by examining the effect of contextual characteristics on state social anxiety, we were confined to only 3 contextual features, which do not encompass the multifaceted nature of real-world social interactions. Future studies should aim to incorporate a broader array of contextual elements (e.g., how familiar is the interaction partner?; \cite{vittengl1998positive}). Additionally, future work that integrates more user self-reported data (such as open-ended text or audio journaling \cite{nepal2024contextual}) surrounding their current social contexts can enhance the richness of our understanding of contextual influences. Finally, our sample was relatively homogeneous on many characteristics (e.g., mostly White, all university students), potentially limiting generalizability of our findings.

%% file: 7_conclusion.tex
\section{Conclusion}

This study explored the feasibility of predicting fluctuations in state social anxiety during virtual social interactions using passive sensing data collected from wearable devices. The findings contribute to a growing body of research suggesting the feasibility of inferring mental health states, such as anxiety, through a combination of biobehavioral markers, contextual, and trait information. The results indicate that integrating multiple modalities enhances predictive performance compared to relying solely on passive biobehavioral indicators. Further, social context characteristics, such as number of interaction partners and timing of the assessment relative to the social interaction (e.g., before, during, or after), emerged as particularly influential in predicting state anxiety, underscoring the importance of collecting and incorporating such data into future models. Additionally, trait measures, which capture broader mental health functioning, were found to enhance model performance (though less so than social context information). Future research exploring state anxiety in naturalistic, face-to-face settings will further refine our understanding of the capabilities of mobile sensing for detecting state anxiety, and set the stage for delivering more personalized interventions.

%% file: Appendix.tex
\appendix
\section{Appendix}

\subsection{Detailed Supplementary Descriptive Statistics on State Anxiety Self-Reports} \label{appendix:descriptive_stats}

This section delves further into the descriptive statistical analysis of our dataset, adding onto results presented in Section \ref{sec:statistical_results}. Before performing feature and machine learning analysis, we examined the proportion and variability of state anxiety self-reports across the sample and within particular individuals. Our analysis informed decisions on how to handle low-variability participants and allowed us to evaluate the appropriateness of the threshold set (self-report > 3) defining anxiety status, thus informative for the subsequent feature and machine learning analysis detailed in Sections \ref{section:r3} and \ref{sec:machine-learning}.

\subsubsection{Participants with Low Variability in Self-Report} \label{app:low_varaibility}

\begin{figure}[h]
    \centering
    \includegraphics[width=0.7\columnwidth]{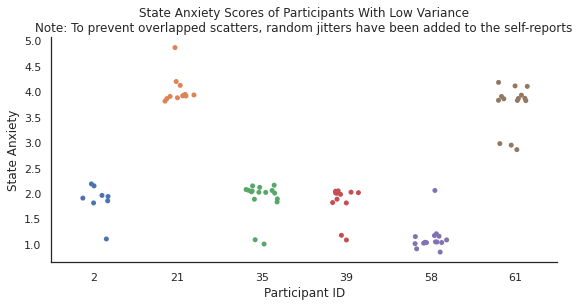}
    \caption{The self-reported scores of the six individual with low variability in self-reports.}
    \label{fig:low-detailed}
\end{figure}

We filtered participants' self-reports to identify those with a standard deviation below 0.5 by observing the distribution of individual state anxiety standard deviations across the sample. This led us to identify six individuals (P2, P21, P35, P39, P58, P61) who exhibited the lowest variability in their responses, such that they consistently reported high or low levels of anxiety. The distributions of these participants' self-reports are depicted in Figure \ref{fig:low-detailed}.

In deciding whether to include or exclude these participants from our sample, we conducted comparative analyses and observed that the key findings remained stable regardless of their inclusion or exclusion. This decision was also informed by the alignment between our observations and the broader patterns in our data. To be specific, SIAS scores and state anxiety means for these six participants were congruent, indicating that participants who reported low trait anxiety also tended to report low state anxiety, and those reporting higher state anxiety tended to report higher state anxiety. Therefore, we concluded that the low variability likely described how these participants truly experienced the social experiences, and we opted to include their data in analyses. As a result, we opted to include these six participants in our analysis in the paper.

\subsubsection{Proportion of High State Anxiety Scores} \label{app:high_state_anxiety}

\begin{figure}[h]
\centering
\includegraphics[width=0.9\columnwidth]{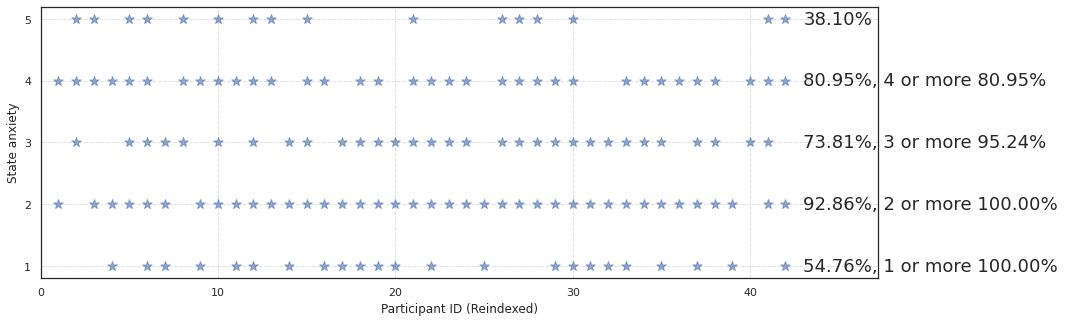}
\caption{Proportion of reported state anxiety levels by participants. On the right side of the figure, the first percentage number indicates the proportion of participants reporting each score at least once; the second number provides cumulative proportions for reporting that score or higher.}
\label{fig:section2-1}
\end{figure}

This section examines the proportion of higher state anxiety scores (4 or 5) across participants. In Figure \ref{fig:section2-1}, we display the range of state anxiety scores each participant reported, with ``stars'' indicating the presence of a particular score. The x-axis features the reindexed participant IDs for clarity, while the y-axis represents the anxiety scores, scaled from 1 to 5. Each ``star'' denotes at least one occurrence of a specific anxiety score reported by a participant during the study.

The analysis shows that every participant reported an anxiety level of at least 2, with more than 95\% reporting scores of 3 or higher. As the anxiety level increases, fewer participants reported higher scores: over 80\% reported a score of 4 or more, while 38\% reported the maximum score of 5. This distribution reinforced our choice of using ``self-report > 3'' as the threshold for defining high anxiety, as it captures a significant proportion of the participants' reported experiences.

\subsection{Experience Length-Independent Time Window Analysis} \label{app:feature_setting}

In addition to the feature extraction methods detailed in the main paper, we explored an alternative strategy to further validate our findings. In addition to the previously employed non-overlapping averaged one-minute time window features, as detailed in Section \ref{sec:feature_extract}, this approach entailed treating the entirety of each experimental session (2, 4, or 6 minutes) as a single, continuous window for feature extraction. We term this approach "single window approach" for the remainder of this appendix. 

Given the variability in session lengths, we reviewed extant literature and found support for the notion that physiological features exhibit relative consistency across time scales. Notably, HRV metrics demonstrate this consistency in ultra-short and short-term contexts\footnote{Malik et al. \cite{malik1996heart} categorized HRV analysis based on 24-hour recordings as \textit{long term}, and analysis up to 5 minutes as \textit{short term}; Castaldo et al. \cite{castaldo2019ultra} defined HRV analysis under 5 minutes as \textit{ultra-short term} for emerging wearable sensing contexts.}, as shown in studies by Castaldo et al. \cite{castaldo2019ultra} and Delaney et al. \cite{delaney2000effects}. Existing findings have supported that variant window sizes should not significantly impact the analysis \cite{munoz2015validity,shaffer2020critical,baek2015reliability}, and also point to the utility of comparing features extracted from different window lengths. Additionally, findings from this comparison may point to alternative ways to handle feature and machine learning analysis in practice. The outcomes of this all-encompassing time window analysis are presented below. 

\subsubsection{Additional Statistical Analysis Results}

In this section, we report statistical analyses using our single window approach. These analyses serve as a parallel comparison to the individual feature-level exploratory analysis detailed in Section \ref{section:r3}. Employing mixed-effects logistic regression, we examine the correlations between state anxiety and physiological features under a different temporal lens.

The results, as shown in Table \ref{table:logistic_raw_value_appendix}, demonstrate significant correlations between various physiological and contextual features with state anxiety levels. Notable findings include increased number of significant associations between state anxiety and HRV features such as SDNN, RMSSD, and SDSD. Also, Accelerometer X-axis Mean was significantly associated with state anxiety. The contextual feature outcomes are consistent with the analysis reported in the main text.

\begin{table}[h]
\centering
\small
\caption{Based on biobehavioral features extracted using our single window approach, mixed-effect logistic regression results for each feature against state anxiety self-reports (raw value > 3 as anxious status vs. non-anxious status). Model Structure: State\_Anxiety \textasciitilde Single\_Feature + (1 + Single\_Feature|Participant\_ID).}
\begin{tabular}{lcccc}
\toprule
\textbf{Feature} & \textbf{Estimate} & \textbf{Std. Error} & \textbf{z value} & \textbf{p value} \\ [0.5ex] 
\midrule
\multicolumn{5}{l}{\textit{\textbf{PPG HRV Features}}} \\
Standard Deviation of NN intervals (SDNN) & -0.5014 & 0.0017 & -289.7 & $<$2e-16*** \\
Root Mean Square of Successive Differences (RMSSD) & -0.6102 & 0.0014 & -437.8 & $<$2e-16*** \\
Standard Deviation of Successive Differences (SDSD) & -0.6071 & 0.0014 & -428.6 & $<$2e-16*** \\
Coefficient of Variation of SD (CVSD) & -0.5863 & 0.2855 & -2.054 & 0.04001* \\
Maximum of NN intervals (MaxNN) & -0.5042 & 0.2506 & -2.012 & 0.04422* \\
Low Frequency Power (LF) & 0.4869 & 0.2134 & 2.281 & 0.02254* \\
Very High Frequency Power (VHF) & 0.3247 & 0.1494 & 2.174 & 0.02969* \\
Natural Log of High Frequency Power (LnHF) & 0.4337 & 0.2118 & 2.048 & 0.0406* \\
Poincaré Plot Standard Deviation 1 (SD1) & -0.6071 & 0.0014 & -428.6 & $<$2e-16*** \\

\midrule
\multicolumn{5}{l}{\textit{\textbf{Accelerometer Feature}}} \\
Accelerometer X-axis Mean & 0.4308 & 0.2143 & 2.011 & 0.04438* \\
\midrule
\multicolumn{5}{l}{\textit{\textbf{Contextual Features}}} \\
Group Size (Dyadic vs. Group)  &  0.7947   &   0.2520  & 3.154 & 0.0016** \\
Interaction Phase - If Anticipatory & 1.3154 & 0.2491 & 5.281 & 1.29e-07*** \\
Interaction Phase - If Post Event & -1.333  &  0.250 & -5.329 & 9.86e-08***\\
\bottomrule
\end{tabular}
\label{table:logistic_raw_value_appendix}
\end{table}

\subsubsection{Additional Machine Learning Analysis Results}

The machine learning models were evaluated under the same configurations described in Section \ref{sec:machine-learning}, while using our single window approach. The performance metrics, summarized in Table \ref{tab:results_with_context_trait_appendix} (compared to Table \ref{tab:results_with_context_trait}), indicate no significant overall change in model performance when using the single window approach.

Specifically, the balanced accuracies for detecting anxiety (self-report > 3) slightly shifted from 69.5\% in the main analysis to 69.0\% in this supplementary analysis. For the detection of extreme anxiety (self-report = 5), the accuracies were closely matched, with 84.5\% in the main analysis compared to 85.2\% in the supplementary analysis. A notable observation was the performance shift between models; XGBoost showed improved performance over the MLP in this setting.

\begin{table}[h]
\caption{\textbf{Performance of machine learning models \textbf{with biobehavioral, trait, and contextual features} using the \textbf{single window approach} with different predictive outcome and data processing configurations.}}
\begin{adjustbox}{width=\textwidth,center}
\small
\centering
\begin{tabular}{lcccccccc}
\toprule
\textbf{Self-Reports} & \multicolumn{4}{c}{\textbf{Anxious vs. Calm (Self-Report $>$ 3)}} & \multicolumn{4}{c}{\textbf{Extremely Anxious (Self-Report = 5)}} \\
\cmidrule(lr){2-5} \cmidrule(lr){6-9}
\textbf{Feature Standardize} & \multicolumn{2}{c}{\textbf{Yes}} & \multicolumn{2}{c}{\textbf{No}}
& \multicolumn{2}{c}{\textbf{Yes}} & \multicolumn{2}{c}{\textbf{No}} \\
\cmidrule(lr){2-3} \cmidrule(lr){4-5} \cmidrule(lr){6-7} \cmidrule(lr){8-9}
\textbf{Model} & \textbf{Accuracy} & \textbf{Macro-F1} & \textbf{Accuracy} & \textbf{Macro-F1} & \textbf{Accuracy} & \textbf{Macro-F1} & \textbf{Accuracy} & \textbf{Macro-F1} \\
\midrule
Baseline (Random Guess) & $0.500\pm0.000$ & $0.500\pm0.000$ & $0.500\pm0.000$ & $0.500\pm0.000$ & $0.500\pm0.000$ & $0.500\pm0.000$ & $0.500\pm0.000$ & $0.500\pm0.000$ \\
\midrule
Gradient Boost & $0.661\pm0.008$ & \textbf{$0.699\pm0.005$} & $0.668\pm0.016$ & \textbf{$0.699\pm0.017$} & \textbf{0.852 $\pm$ 0.004} & \textbf{0.904 $\pm$ 0.004} & 0.828 $\pm$ 0.002 & 0.877 $\pm$ 0.001 \\
XGBoost & \textbf{0.690} $\pm$ \textbf{0.000} & \textbf{0.717} $\pm$ \textbf{0.000} & \textbf{0.698} $\pm$ \textbf{0.000} & \textbf{0.737} $\pm$ \textbf{0.000} & 0.843 $\pm$ 0.000 & 0.886 $\pm$ 0.000 & \textbf{0.843} $\pm$ \textbf{0.000} & 0.884 $\pm$ 0.000 \\
Random Forest & $0.614\pm0.023$ & $0.650\pm0.026$ & $0.608\pm0.012$ & $0.645\pm0.018$ & 0.850 $\pm$ 0.001 & 0.899 $\pm$ 0.001 & 0.833 $\pm$ 0.005 & \textbf{0.885} $\pm$ \textbf{0.003} \\
Decision Tree & $0.614\pm0.015$ & $0.631\pm0.022$ & $0.578\pm0.013$ & $0.627\pm0.016$ & 0.841 $\pm$ 0.003 & 0.896 $\pm$ 0.002 & 0.782 $\pm$ 0.012 & 0.845 $\pm$ 0.010 \\
Multilayer Perceptron & $0.669\pm0.023$ & $0.681\pm0.025$ & $0.657\pm0.025$ & $0.676\pm0.037$ & 0.839 $\pm$ 0.005 & 0.889 $\pm$ 0.007 & 0.823 $\pm$ 0.005 & 0.874 $\pm$ 0.003 \\
Logistic Regression & $0.634\pm0.000$ & $0.649\pm0.000$ & $0.679\pm0.000$ & $0.713\pm0.000$ & 0.844 $\pm$ 0.000 & 0.876 $\pm$ 0.000 & 0.833 $\pm$ 0.000 & 0.876 $\pm$ 0.000 \\
SVM Classifier & $0.623\pm0.000$ & $0.668\pm0.000$ & $0.657\pm0.000$ & $0.689\pm0.000$ & 0.845 $\pm$ 0.000 & 0.888 $\pm$ 0.000 & 0.835 $\pm$ 0.000 & 0.878 $\pm$ 0.000 \\
K-Nearest Neighbors & $0.622\pm0.000$ & $0.661\pm0.000$ & $0.634\pm0.000$ & $0.655\pm0.000$ & 0.840 $\pm$ 0.000 & 0.884 $\pm$ 0.000 & 0.829 $\pm$ 0.000 & 0.877 $\pm$ 0.000 \\
\bottomrule
\end{tabular}
\end{adjustbox}
\label{tab:results_with_context_trait_appendix}
\end{table}

\subsection{Analytical Results Across Different Anxiety Outcome Operationalizations} \label{app:outcome_three}

We explored different operationalizations of state anxiety status beyond the self-report > 3 threshold reported in the main text. Our analysis included two alternative operationalizations: 1) within-person mean-based, where state anxiety status is defined by self-reported measures surpassing the individual-level within-person mean, and 2) baseline-based, where state anxiety status is determined by current self-reported state anxiety exceeding the baseline established before the anticipatory anxiety phase. These outcomes are presented in Table \ref{tab:model_performance_appendix} and can be compared to the results in Table \ref{tab:results_with_context_trait}, which were derived using the same feature set and model configurations but varied in defining state anxiety status.

\begin{table}[h]
\caption{\textbf{Model performance under two predictive outcome configurations:} 1) self-reported measures exceeding individual-level within-person mean as state anxiety status and 2) current self-reported state anxiety exceeding baseline before anticipatory anxiety phase, using the same features and model configurations as in Table \ref{tab:results_with_context_trait}.}
\begin{adjustbox}{width=\textwidth,center}
\small
\centering
\begin{tabular}{lcccccccc}
\toprule
\textbf{Self-Reports} & \multicolumn{4}{c}{\textbf{Operationalization II. Within-Person Elevations $>$ 0}} & \multicolumn{4}{c}{\textbf{Operationalization III. Compared to the Baseline}} \\
\cmidrule(lr){2-5} \cmidrule(lr){6-9}
\textbf{Feature} & \multicolumn{2}{c}{\textbf{Person-Standardized: True}} & \multicolumn{2}{c}{\textbf{Person-Standardized: False}}
& \multicolumn{2}{c}{\textbf{Person-Standardized: True}} & \multicolumn{2}{c}{\textbf{Person-Standardized: False}} \\
\cmidrule(lr){2-3} \cmidrule(lr){4-5} \cmidrule(lr){6-7} \cmidrule(lr){8-9}
\textbf{Model} & \textbf{Accuracy} & \textbf{Macro-F1} & \textbf{Accuracy} & \textbf{Macro-F1} & \textbf{Accuracy} & \textbf{Macro-F1} & \textbf{Accuracy} & \textbf{Macro-F1} \\
\midrule
Baseline (Random Guess) & $0.500\pm0.000$ & $0.500\pm0.000$ & $0.500\pm0.000$ & $0.500\pm0.000$ & $0.500\pm0.000$ & $0.500\pm0.000$ & $0.500\pm0.000$ & $0.500\pm0.000$ \\
\midrule
Gradient Boost & \textbf{0.673} $\pm$ \textbf{0.013} & \textbf{0.649} $\pm$ \textbf{0.015} & $0.648\pm0.035$ & $0.632\pm0.040$ & $0.491 \pm 0.016$ & $0.466 \pm 0.027$ & $0.435 \pm 0.020$ & $0.451 \pm 0.011$ \\
XGBoost & $0.622\pm0.000$ & $0.627\pm0.000$ & \textbf{0.688} $\pm$ \textbf{0.000} & \textbf{0.653} $\pm$ \textbf{0.000} & $0.501 \pm 0.000$ & $0.492 \pm 0.000$ & $0.458 \pm 0.000$ & $0.527 \pm 0.000$ \\
Random Forest & $0.631\pm0.043$ & $0.614\pm0.041$ & $0.593\pm0.025$ & $0.570\pm0.032$ & $0.485 \pm 0.028$ & $0.473 \pm 0.031$ & $0.495 \pm 0.018$ & $0.540 \pm 0.030$ \\
Decision Tree & $0.640\pm0.008$ & $0.628\pm0.014$ & $0.625\pm0.031$ & $0.600\pm0.022$ & $0.466 \pm 0.028$ & $0.522 \pm 0.018$ & $0.461 \pm 0.046$ & $0.444 \pm 0.028$ \\
Multilayer Perceptron & $0.598\pm0.021$ & $0.577\pm0.012$ & $0.637\pm0.023$ & $0.592\pm0.017$ & $0.485 \pm 0.018$ & $0.445 \pm 0.026$ & $ 0.480 \pm 0.020$ & $ 0.455 \pm 0.025$ \\
Logistic Regression & $0.634\pm0.000$ & $0.600\pm0.000$ & $0.633\pm0.000$ & $0.595\pm0.000$ & $0.518 \pm 0.000$ & $0.542 \pm 0.000$ & $ 0.515 \pm 0.000$ & $ 0.540 \pm 0.000$ \\
SVM Classifier & $0.651\pm0.000$ & $0.626\pm0.000$ & $0.593\pm0.000$ & $0.563\pm0.000$ & $0.529 \pm 0.000$ & $0.535 \pm 0.000$ & $ 0.525 \pm 0.000$ & $ 0.530 \pm 0.000$ \\
K-Nearest Neighbors & $0.589\pm0.000$ & $0.581\pm0.000$ & $0.624\pm0.000$ & $0.593\pm0.000$ & $0.456 \pm 0.000$ & $0.452 \pm 0.000$ & $ 0.460 \pm 0.000$ & $ 0.455 \pm 0.000$ \\
\bottomrule
\end{tabular}
\end{adjustbox}
\label{tab:model_performance_appendix}
\end{table}

In the `Within-Person Elevations > 0' setting, the decrease in model accuracies compared to `Raw Self-Report > 3' suggests weaker identification outcomes of detecting state anxiety, indicating that individualized (i.e., within-person) benchmarks might complicate consistent anxiety state classification. Additionally, the baseline-based outcome setting revealed limited predictive capability, underscoring the challenge in utilizing baseline comparisons for anxiety status determination.